\documentclass[final,5p,times,twocolumn]{elsarticle}
\usepackage{tabularx}
\usepackage{ulem}
\DeclareGraphicsExtensions{.pdf,.jpg}
\usepackage[cmex10]{amsmath}

\makeatletter
\def\vec@style{\relax} 
\def\vec#1{\relax\ifmmode\mathchoice
{\mbox{\boldmath$\vec@style\displaystyle#1$}}
{\mbox{\boldmath$\vec@style\textstyle#1$}}
{\mbox{\boldmath$\vec@style\scriptstyle#1$}}
{\mbox{\boldmath$\vec@style\scriptscriptstyle#1$}}\else
\hbox{\boldmath$\vec@style\textstyle#1$}\fi}

\def\mat@style{\sf} 

\def\mat#1{\relax\ifmmode\mathchoice
{\mbox{\boldmath$\mat@style\displaystyle#1$}}
{\mbox{\boldmath$\mat@style\textstyle#1$}}
{\mbox{\boldmath$\mat@style\scriptstyle#1$}}
{\mbox{\boldmath$\mat@style\scriptscriptstyle#1$}}\else
\hbox{\boldmath$\mat@style\textstyle#1$}\fi}
\makeatother
\newcommand{\up}[1]{{\text{#1}}}
\newcommand{\fracd}[2]{\frac{\displaystyle #1}{\displaystyle #2}}

\begin{document}

\begin{frontmatter}

\title{Light-Trapping Enhanced Thin-Film III-V Quantum Dot Solar Cells Fabricated by Epitaxial Lift-Off}

\author[polito]{F. Cappelluti\corref{cor1}}
\ead{federica.cappelluti@polito.it}
\author[ucl]{D. Kim}
\author[rbu]{M. van Eerden}
\author[polito]{A. P. C\'edola}
\author[tut]{T. Aho}
\author[tf2]{G. Bissels}
\author[polito]{F. Elsehrawy}
\author[ucl]{J. Wu}
\author[ucl]{H. Liu}
\ead{huiyun.liu@ucl.ac.uk}
\author[rbu]{P. Mulder}
\author[rbu]{G. Bauhuis}
\author[rbu]{J. Schermer}
\ead{j.schermer@science.ru.nl}
\author[tut]{T. Niemi}
\author[tut]{M. Guina}
\ead{mircea.guina@tut.fi}

\cortext[cor1]{Corresponding author}

\address[polito]{Department of Electronics and Telecommunications, Politecnico di Torino, \\ Corso Duca degli Abruzzi  24, 10129 Torino, Italy}
\address[ucl]{Department of Electronic and Electrical Engineering, University College London, Torrington Place, London WC1E 7JE, United Kingdom}
\address[rbu]{Institute for Molecules and Materials, Radboud University, Toernooiveld 1, 6525 ED Nijmegen, The Netherlands}
\address[tut]{Optoelectronics Research Centre, Photonics Laboratory, Tampere University of Technology, P.O. Box 692, FIN-33101 Tampere, Finland}
\address[tf2]{tf2 devices B.V., Heyendaalseweg 135, 6525 AJ Nijmegen, The Netherlands}

\begin{abstract}
We report thin-film InAs/GaAs quantum dot (QD) solar cells with $n-i-p^{+}$ deep junction structure and planar back reflector fabricated by epitaxial lift-off (ELO) of full 3-inch wafers. External quantum efficiency measurements demonstrate twofold enhancement of the QD photocurrent in the ELO QD cell compared to the wafer-based QD cell. In the GaAs wavelength range, the ELO QD cell perfectly preserves the current collection efficiency of the baseline single-junction ELO cell. We demonstrate by full-wave optical simulations that integrating a micro-patterned diffraction grating in the ELO cell rearside provides more than tenfold enhancement of the near-infrared light harvesting by QDs. 
Experimental results are thoroughly discussed with the help of physics-based simulations to single out the impact of QD dynamics and defects on the cell photovoltaic behavior. It is demonstrated that non radiative recombination in the QD stack is the bottleneck for the open circuit voltage ($V_\up{oc}$) of the reported devices. More important, our theoretical calculations demonstrate that the $V_\up{oc}$ offest of 0.3 V from the QD ground state identified by \emph{Tanabe et al., 2012}, from a collection of experimental data of high quality III-V QD solar cells is a reliable - albeit conservative - metric to gauge the attainable $V_\up{oc}$ and to quantify the scope for improvement by reducing non radiative recombination. Provided that material quality issues are solved, we demonstrate - by transport and rigorous electromagnetic simulations - that light-trapping enhanced thin-film cells with twenty InAs/GaAs QD layers reach efficiency higher than 28\% under unconcentrated light, ambient temperature. If photon recycling can be fully exploited,  30\%  efficiency is deemed to be feasible.
\end{abstract}


\end{frontmatter}


\section{Introduction}
\label{intro}
Nanostructured absorbers based on quantum-dots (QD) provide tunable sub-bandgap transitions to enhance the infrared photoresponse of single-junction solar cells \cite{2001Aroutiounian_JAP} and to improve current matching in multijunction cells \cite{2014Kerestes_PPRA,2016Ho_PPRA}. Also, they offer a promising path towards the development of novel photovoltaics concepts, beyond the Shockley-Queisser (SQ) limit, such as the intermediate band (IB) solar cell \cite{1997Luque_PRL,2006Green_Third}. By leveraging on second photon absorption or hot phonons, power conversion efficiency well above the SQ limit is theoretically achievable in QD enhanced single-junction cells \cite{2006Green_Third,2001Luque_TED,2007Wei_APL}.

The most widely investigated structure of QD solar cells (QDSCs) exploits a stack of InAs/GaAs QD layers embedded in a single-junction GaAs solar cell \cite{2001Aroutiounian_JAP}. A clear enhancement of the infrared spectral response is usually observed in these devices, but the maximum demonstrated efficiency (18.7\% at 1 sun) \cite{2012Tanabe_APL} lags well behind that one of state-of-art GaAs single-junction cells (28.8\% at 1 sun \cite{2011Kayes_PVSC,2017Green_SolarTable}), not to mention the gap with respect to the efficiency predicted by the IB theory ($\approx 36\%$ for the InAs/GaAs material system at 1 sun). The large discrepancy between demonstrated and theoretical efficiency is often ascribed to the fact that reported devices do not work in the IB operating regime, because at room temperature - and especially under unconcentrated light - the second photon absorption is irrelevant compared to the thermally activated escape. However, for QD cells operating in such thermally-limited regime, a conservative reference value for the attainable efficiency is given by a  conventional single-junction cell with bandgap comparable to the optical transition energy of the QD ground state: assuming a band gap of 1 eV  (representative of the InAs/GaAs QD system \cite{2012Tanabe_APL}), the SQ limit efficiency is above 30\% under 1 sun, a value remarkably higher than the demonstrated 18.7\%.

The issue arises primarily from the weak increase of the short circuit current provided by the inclusion of QDs, owing to the small optical absorption cross-section of the sub-bandgap transitions and the reduced QD volume fraction within the absorbing region. On the other hand, one of the conditions to attain the single-gap SQ limit is complete interband absorptivity, while approaching the IB operating regime requires very high - and similar - interband and intraband photon absorption \cite{2012Sakamoto_JAP,2014Mellor_SEMSC,2015Okada_APR}. Detailed balance calculations in \cite{2012Sakamoto_JAP} show that high efficiency InAs/GaAs QDSCs operating in the IB regime require a total QD density  larger than $5\times 10^{13}$ cm$^{-2}$, but present QDSCs usually have a few tens of layers and in-plane density about $5\times 10^{10}$ cm$^{-2}$. Thus, significant research efforts are being carried out to increase the areal density of III-V QDs \cite{2010Zhou_APL,2011Fujita_PVSC,2013Tutu_APL2,2016Sameshima_APE} and the number of QD layers \cite{2011Akahane_PPSa,2012Sugaya_EES} without compromising crystal quality. In-plane density up to $10^{12}$ cm$^{-2}$ \cite{2016Sameshima_APE} and number of QD stacks up to 400 \cite{2012Sugaya_EES}  have been reported, but QDSCs with both high number of QD layers and high in-plane QD density have not yet been demonstrated. Moreover, present QDSCs often suffer of severe open circuit voltage ($V_\up{oc}$) degradation, which tends to get worse as the density or the number of layers are increased \cite{2012Sugaya_EES}. In general, achieving high crystal quality is one of the major technological challenges within QDSC research, since QD-growth induced defects markedly impair the $V_\up{oc}$ \cite{2010Guimard_APL,2016Cappelluti_SEMSC}. In high-quality QDSCs operating in the thermally-limited regime, experimental \cite{2010Guimard_APL,2012Tanabe_APL} and theoretical \cite{2010Jolley_APL,2013Gioannini_JPV} works show that the maximum attainable $V_\up{oc}$ is linearly correlated with the energy band gap of the QD ground state. $V_\up{oc}$ approaching 1 V has been demonstrated only in $10\times$ and $40\times$ QD layer cells (with shallow QDs and in-plane density well below $10^{11}$ cm$^{-2}$) by implementing complex strain compensation techniques during the epitaxial growth \cite{2011Bailey_APL,2012Bailey_JPV,2016Smith_PVSC}.

A promising alternative (and even somewhat complementary) path to effectively enhance QD photogeneration is offered by light management schemes that can be implemented within a thin-film solar cell architecture \cite{2015Welser_book,2016Smith_PVSC,2016Cappelluti_PVSC}. Thin-film III-V technologies based on epitaxial lift-off (ELO) \cite{2008Tatavarti_PVSC,2009Bauhuis_SEMSC,2011Kayes_PVSC} are among the most promising approaches in view of the remarkable reduction of mass and cost (because ELO makes possible wafer reuse), and flexibility. Moreover, photon trapping and recycling enabled by the thin-film design have been proven to be essential to push the efficiency of III-V single-junction cells towards the SQ limit \cite{1991Lush_SC,2012Miller_JPV,2013Wang_JPV}, and are in fact at the root of the world-record 28.8\% efficiency held by an ELO thin-film GaAs cell with planar mirrored rear surface \cite{2011Kayes_PVSC,2017Green_SolarTable}. ELO thin-film InAs/GaAs QD cells with planar rear mirror were first reported in \cite{2013Bennett_APL,2014Sogabe2014_APL}. More recently, ELO thin-film cells with textured back surface reflectors have been reported in \cite{2016Smith_PVSC}, demonstrating a 30\% increase of QD current contribution compared to a cell with planar reflector. Effective ligh-trapping by back side periodic grooves has been demonstrated in multiple quantum well solar cells in \cite{2015Inoue_JPV}, attaining a fivefold increase of the sub-bandgap optical path length.

In this work, we report thin-film InAs/GaAs QD solar cells fabricated by epitaxial lift-off of full 3-inch wafers. Nearly doubled QD photocurrent is demonstrated in the thin-film ELO QD cell with planar gold mirror with respect to a baseline wafer-based QD cell. Moreover, collection efficiency within the GaAs wavelength range is perfectly preserved in the ELO QD cell compared to the ELO baseline single-junction cell. Solar cell performance are analyzed with the support of device-level physics-based simulations, with the aim at providing an assessment of the needs in terms of material optimization and QD photogeneration enhancement to attain high efficiency InAs/GaAs QDSCs. The practical implementation of photonic structures able to provide the desired high QD sub-bandgap harvesting is discussed based on rigorous wave-optics simulations of broadband antireflection coatings and diffraction gratings that can be fabricated by patterning the front and rear surface of the thin-film cell.

The rest of the paper is organized as follows. Sec.\,\ref{mat} summarizes the details of device fabrication and characterization, while Sec.\,\ref{model} describes the theoretical background and numerical tools used for the analysis of experimental results and the design of light-trapping enhanced cells. The experimental results of wafer-based and thin-film ELO cells are discussed in Sec.\,\ref{sec_results_1} and Sec.\,\ref{sec_results_2}, respectively, with the help of physics-based simulations. Finally, in Sec.\,\ref{sec_results_3}, the performance of thin-film QD cells integrating photonic structures are discussed based on transport and full-wave electromagnetic simulations.

\section{Material and Methods}
\label{mat}
\begin{figure}
\centerline{\includegraphics[width=.9\columnwidth]{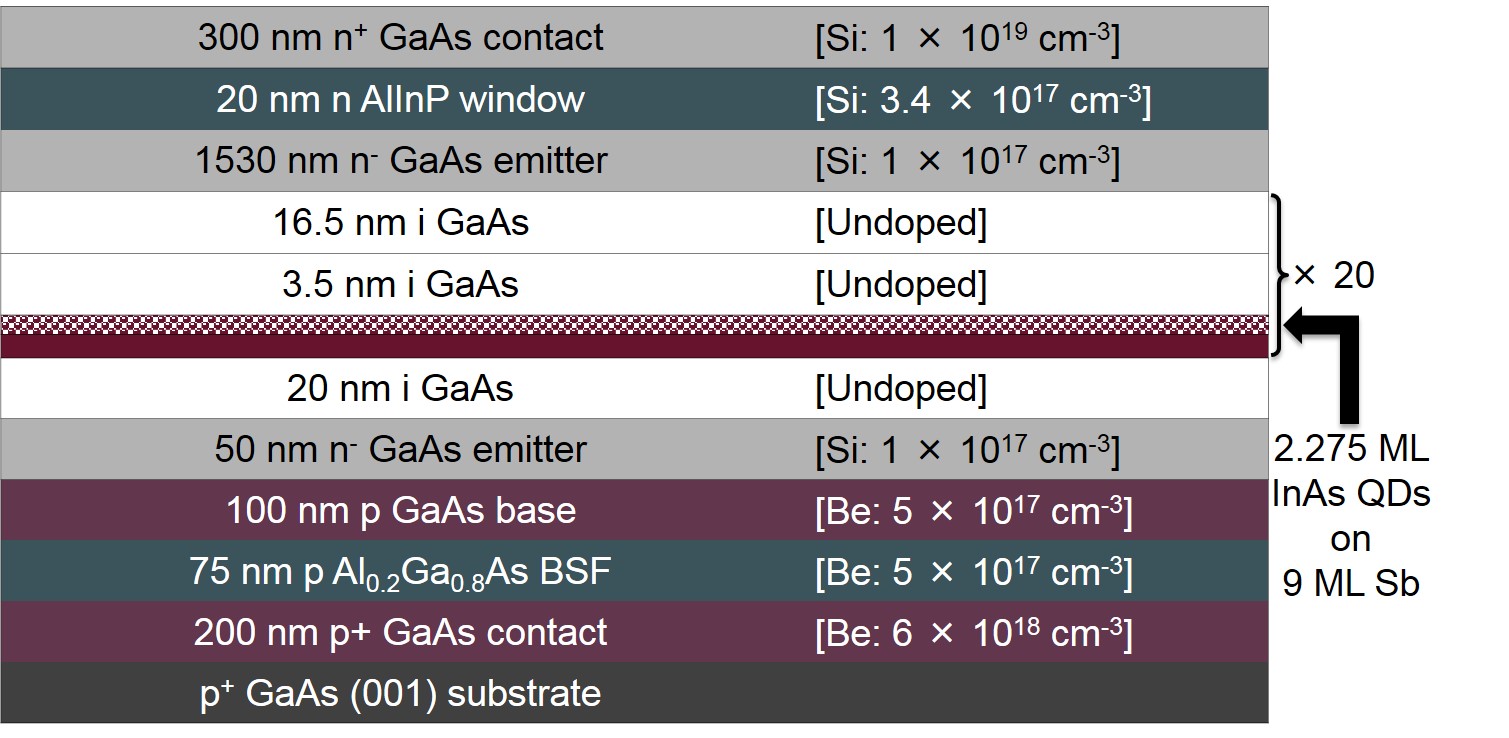}}
\vskip .5cm
\centerline{\includegraphics[width=.9\columnwidth]{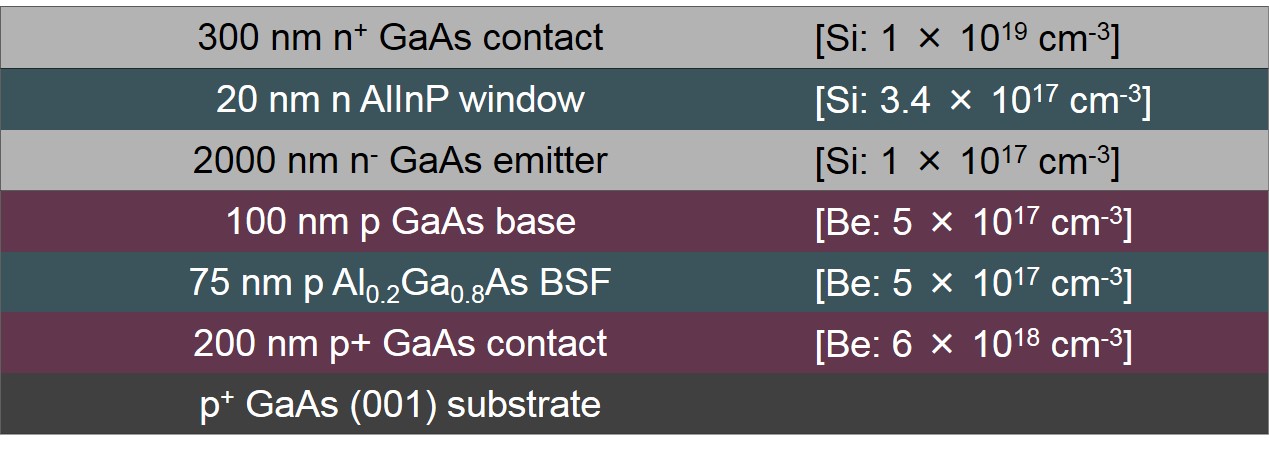}}
\caption{\label{Fig1} Sketch of the epilayer structure of the 20 $\times$ InAs/GaAs QD layers  (top) and regular GaAs (bottom) solar cells. In the $50\times$ QD cell (not shown) the topmost part of the emitter has a reduced thickness of 930 nm to keep the whole emitter (topmost doped region + intrinsic QD stack + bottom emitter region) thickness to $\approx 2$ $\mu$m.}
\end{figure}

QD and regular GaAs cells with a deep junction design with lightly $n$-doped emitter and thin $p$-doped base are studied. Several batches of wafers, both for wafer-based processing and for ELO thin-film processing, were grown by molecular beam epitaxy (MBE). Fig.\,\ref{Fig1} reports the detailed epilayer structure for the wafer-based InAs/GaAs QD and regular GaAs solar cells. The QDSCs exploit high in-plane density (over $8 \times 10^{10}$ cm$^{-2}$) InAs/GaAs QD layers fabricated through the Sb-mediated QD growth technique, following the method already demonstrated in \cite{2013Tutu_APL2}.  The QD periodic stack is placed in the bottom part of the emitter and uses 20 nm thick spacer layers of intrinsic GaAs. Samples with 20$\times$ and $50\times$ QD layers were fabricated. All the cells - GaAs-only and QD-based - have a total emitter thickness of $\approx2\,\mu$m. For the thin-film configuration the epilayer structures were identical, except for the window and back surface field layers which were made by InGaP, and the inclusion of \emph{ad-hoc} release and etch stop layers for the ELO processing. The ELO thin-film processing was performed as described in detail in \cite{2006Schermer_TSF}. Because the cell structures were only produced for mutual comparison no efforts were made to optimize the front grid contact coverage or to apply an ARC to minimize the reflection from the front surface. The thin-film cells exploit a planar gold mirror on the backside.

$I-V$ characterization of the solar cells was performed using an ABET Technologies Sun 2000 Class A solar simulator, which provides homogeneous illumination over a 100 $\times$ 100  mm$^2$ area. An Ushio 550W Xenon short arc lamp is used to approximate the AM1.5 spectrum. The setup is equipped with a Keithley 2600 sourcemeter and data acquisition is performed using ReRa Tracer3 software. The solar cells are kept at 25 $^\mathrm{o}$C during measurement by water cooling. External Quantum Efficiency (EQE) measurements were performed with a ReRa SpeQuest Quantum Efficiency system. Data acquisition is performed using ReRa Photor 3.1 software. The system uses both a Xenon and halogen light source to access all wavelengths present in the solar spectrum. A monochromator is used to generate quasi-monochromatic light and a chopper for intensity modulation. This generates a test light of variable wavelength while a continuous bias light is used to put the cell under test in operating conditions. Additionally infrared electroluminescence images were captured using a set-up consisting of a Thorlabs CMOS camera with NIR sensor and a home-build power supply able to inject a stable current to the cells over a large range.

\section{Theory and Calculations}
\label{model}
\subsection{Electrical model}
The photovoltaic performance of InAs/GaAs QDSCs are discussed based on numerical simulations which take into account the transfer and transport processes involving carriers in the confined states introduced by the nanostructured material and carriers in the extended states of the host bulk material \cite{2013Gioannini_JPV,2016Cappelluti_SEMSC}.

\begin{figure}[ht]
\centerline{\includegraphics[width=.5\columnwidth]{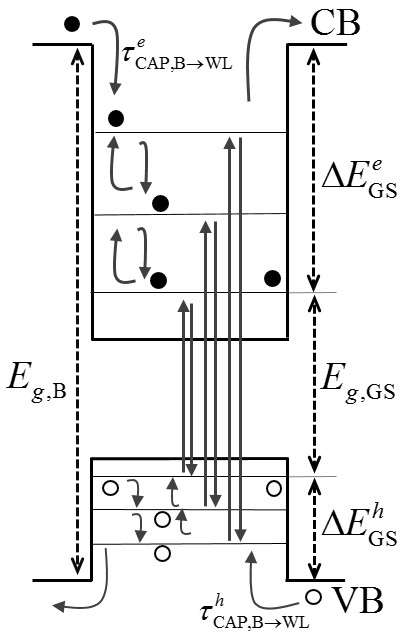}}
\caption{\label{Fig_QD} Sketch of the QD sub-band electronic structure highlighting the interband and intersubband carrier processes taken into account in the model. $E_{g,\up{B}}$, $E_{g,\up{GS}}$ indicate the GaAs and GS bandgap energy, respectively. $\Delta E^e_{\up{GS}}$, $\Delta E^h_{\up{GS}}$ the confinement energy for electrons and holes in the GS.}
\end{figure}

\begin{figure}
\centerline{\includegraphics[width=1\columnwidth]{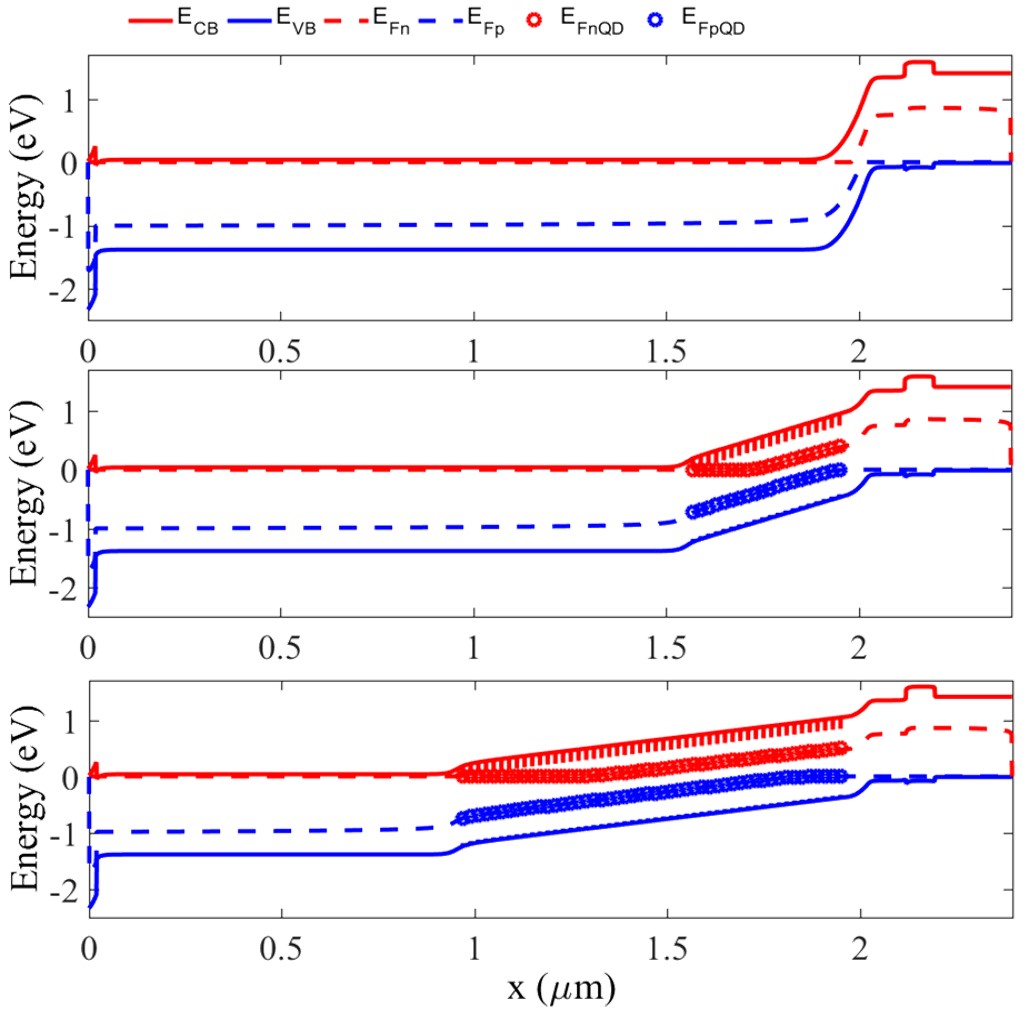}}
\caption{\label{Fig_EBG} Energy band diagram under short circuit condition (1 sun AM1.5G, $T=300$ K) for the regular GaAs cell (top), $20\times$ QD cell (middle), and $50\times$ QD cell (bottom) under study.}
\end{figure}

The three-dimensional confinement of carriers in the QDs is modeled by three discrete energy levels in the conduction and valence band: ground state (GS), excited state (ES), and wetting layer (WL), as sketched in Fig.\,\ref{Fig_QD}. Consider also the energy band diagram of the QDSCs shown in Fig.\,\ref{Fig_EBG}: At each QD layer, several interband and intraband charge transfer processes are possible, including photogeneration, radiative recombination, capture of free carriers in the QD confined states or conversely the excitation of confined carriers from the bound states to the continuum bands, intersubband relaxation and excitation. These processes are described by a set of spatially-resolved (along the QD stack) rate equations self-consistently solved with drift-diffusion and Poisson equations. The detailed model formulation can be found in \cite{2013Gioannini_JPV}. For the sake of the following discussion, it is sufficient to remind that the presented simulations neglect second photon absorption and field-assisted tunneling, i.e. intraband excitation is thermally activated only. Thus, escape time constants are derived from capture and relaxation scattering times based on the argument of detailed balance at thermal equilibrium \cite{2016Cappelluti_SEMSC}.


This modeling approach allows to single out the impact on the cell photovoltaic behavior of different and interplaying physical mechanisms, such as QD kinetics \cite{2013Gioannini_JPV,2014Gioannini_IET,2016Cappelluti_IET}, non radiative recombination \cite{2016Cappelluti_SEMSC}, and photon losses (by proper coupling with a suitable optical model) and thus it is very well suited for the interpretation of experimental results and the investigation of novel design solutions.

QD electronic structure, optical properties, and scattering rates for the charge transfer mechanisms can be derived from quantum-mechanical models or - as done in this work - from experimental data. QD parameters in terms of energy confinement and optical cross-section have been extracted from photoluminescence measurements of prototype samples with one QD stack and from the long-wavelength EQE spectra of the wafer based cells reported in Sec.\ref{sec_results_1}. In particular, the bandgap of GS, ES, and WL states is set to 1.17, 1.25, and 1.37 eV, respectively. The electron and hole confinement energy in the QD levels (e.g.,  for the GS level, $\Delta E^e_{\up{GS}}$ and $\Delta E^h_{\up{GS}}$ in Fig.\,\ref{Fig_QD}) is estimated by assuming that 80\% of the difference between the GaAs band gap ($E_{g,\up{B}}=1.424$ eV) and the QD level energy gap is allocated to the conduction band \cite{2000Williamson_PRB,2016Cappelluti_IET}. Since the thermally-dominated escape rate depends on the confinement energy, this implies that holes have a markedly faster dynamics than electrons. In this situation, numerical simulations \cite{2014Gioannini_IET} demonstrate that at short circuit radiative recombination through the QDs is inhibited, and carrier collection is limited by transport and non radiative recombination of carriers in the barrier. Thus, in an optimized cell, almost all the QD photogenerated carriers are collected at the cell contacts and give a short circuit current contribution that adds up to that one provided by the host material. According to the discussion in \cite{2013Gioannini_JPV}, carrier capture and relaxation times between the QD states range on a scale of 100 fs for holes and 1 ps for electrons, while 1 ns radiative lifetime is used for all the QD levels. In the QD layers, the optical absorption associated to each interband transition is modeled by a Gaussian function with absorption peaks on the order of $10^3$ cm$^{-1}$ for GS and ES states, and $10^4$ cm$^{-1}$ for WL state, assuming an equivalent QD layer thickness of 4 nm and areal density of $8\times10^{10}$ cm$^{-2}$. The full list of QD and bulk material parameters used in this study is summarized in the Supplementary Material.

\subsection{Optical modeling of wafer-based and planar thin-film cells}
In this work, the calculation of the spatially resolved photogeneration rate has been improved with respect to the simple Lambert-Beer approach used in \cite{2013Gioannini_JPV}, by introducing an electromagnetic model which takes into account interference effects. A detailed optical model is in fact essential to ensure an accurate fitting of the measured EQE spectra analysed in Sec.\,\ref{sec_results_1} and Sec.\,\ref{sec_results_2}, and thus a reliable estimation of relevant - technology dependent - material parameters such as carriers lifetime. The optical field distribution across the multilayer structure and the corresponding photogeneration rate are calculated according to a scattering matrix formalism for coherent multilayers \cite{1999Petterson_JAP}. Taking advantage of the spectral separation of the QD and GaAs optical absorption, the QD layers are modeled as an effective medium whose absorption coefficient results from the linear superposition of the GaAs optical absorption and of the below-bandgap absorption induced by QDs as derived from the fitting of the EQE measurement. The real part of the refractive index of the QD layers is then calculated through Kramers-Kronig relation. The complex refractive index of bulk materials is extracted from the database in \cite{SOPRA}.

\subsection{Optical modeling of textured thin-film cells}
\begin{figure}
\centerline{\includegraphics[width=1\columnwidth]{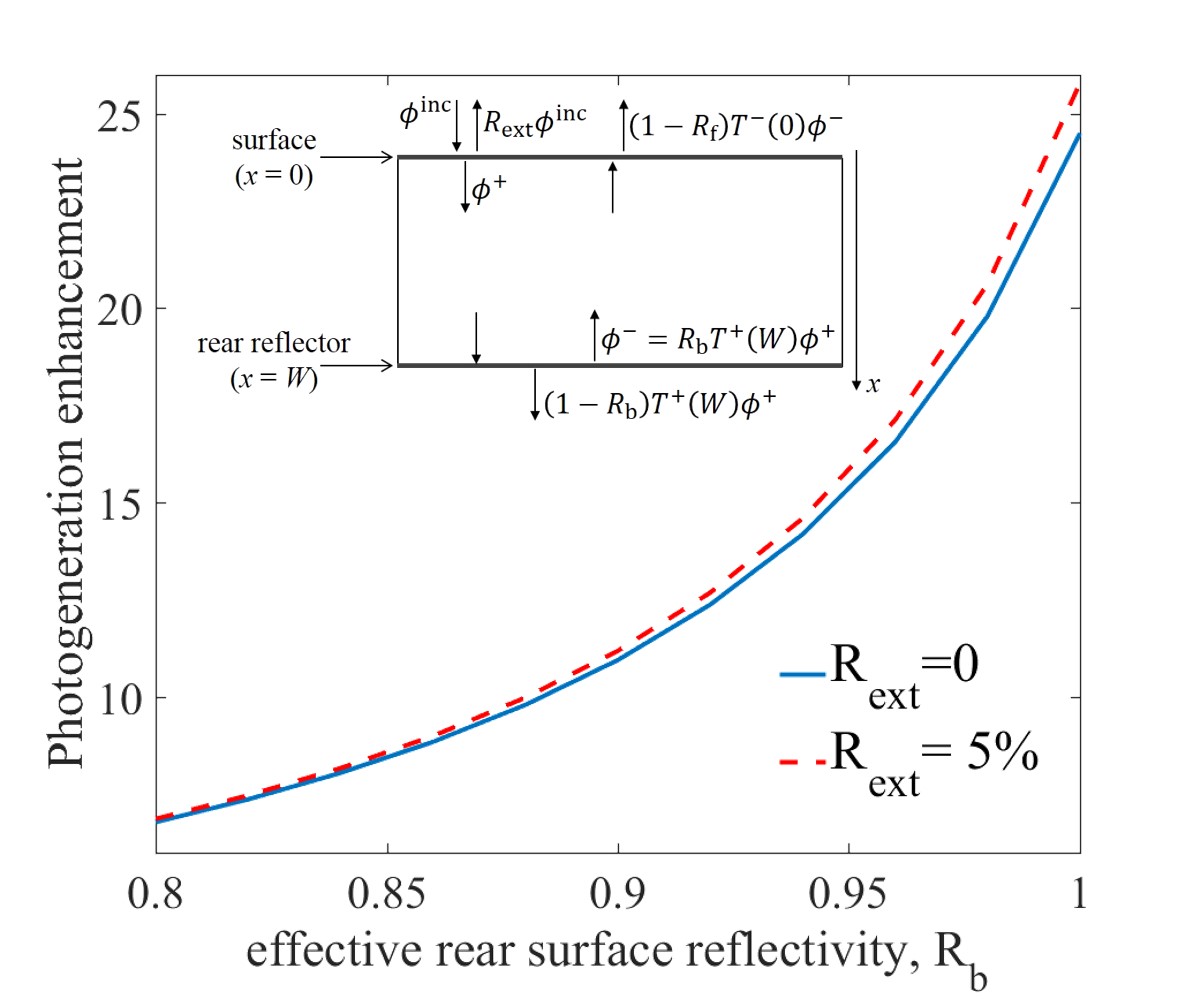}}
\caption{\label{Fig_LT} Enhancement of the photogeneration rate in the thin-film light trapping cell - in the limit of weak absorption ($\alpha W \rightarrow 0$) - as a function of the effective rear surface reflectance according to the incoherent multiple-reflection Lambertian model ($R_\up{f}=1-(1-R_\up{ext})/n^2$) depicted in the inset, for the ideal case of $R_\up{ext}=0$ and the more realistic assumption of $R_\up{ext}=5\%$.}
\end{figure}

Light-trapping in the textured thin-film cells analyzed in Sec.\,\ref{sec_results_3} is modelled by assuming multiple incoherent reflections between the front and rear surfaces characterized by the reflectance $R_\up{f}$ and $R_\up{b}$, respectively. A schematic illustration of the model is shown in the inset of Fig.\,\ref{Fig_LT}. $R_\up{b}$ is intended as the effective reflectance of the cell rear surface, which models e.g. the cumulative effect of the textured grating, planarizing polymer, and metal mirror realized at the backside of the cell. Under the assumption of Lambertian light trapping \cite{1988Gee_PIEEE,2002Green_PPRA}, $R_\up{b}$ is assumed to be angular independent, i.e. the rearside surface works as an ideal diffuse reflector, and the top internal reflectance is given as  $R_\up{f}=1-(1-R_\up{ext})/n^2$, $n$ being the semiconductor refractive index and $R_\up{ext}$ the reflectance at the illuminated surface.
The optical energy flux (W/cm$^2$) in the cell results as the combination of downward and upward propagating fluxes:

\begin{equation}
\label{eq}
\phi(x)=\phi^\up{inc}\left(1-R_\up{ext}\right)\fracd{T^+(x)+R_\up{b}T^+(W)T^-(x)}{1-R_\up{b}R_\up{f}T^+(W)T^-(0)}
\end{equation}

where $\phi^\up{inc}$ is the incident solar flux and $T^{+}$ and $T^{-}$ are the downward and upward transmittance, respectively, calculated assuming perpendicular propagation, i.e. $T^+(W)=T^-(0)=\mathrm{exp}(-\alpha W)$. This, as shown in Fig.\ref{Fig_LT}, leads the maximum achievable enhancement of the cell absorbance (and corresponding photogeneration rate) to approach $2n^2$ ($\approx 25$ for GaAs) in the weak absorption limit, i.e. $\alpha W \rightarrow 0$. As an example, in the $20\times$ QD cell, the weakly absorbing GS state may approach the $2n^2$ limit \cite{2016Cappelluti_PVSC}. In a truly 3D geometry, taking into account the longer path length of oblique rays, and averaging over the angle of propagation, a further factor of 2 is found, yielding the well-known $4n^2$ Lambertian limit \cite{1982Yablo_TED}. In this sense, the presented simulations may be considered somewhat conservative with respect to the $4n^2$ Lambertian limit. Eq.\,(1) also holds in the strong absorption limit ($\alpha W>>1$), where it reduces to $\phi(x)=\phi^\up{inc}\left(1-R_\up{ext}\right)\mathrm{exp}(-\alpha W)$. For $R_\up{f}=0$, eq.\,(1) describes - in the limit of incoherent reflections - a planar cell with rearside mirror with reflectivity $R_\up{b}$.

\begin{figure}[ht]
\centerline{\includegraphics[width=.9\columnwidth]{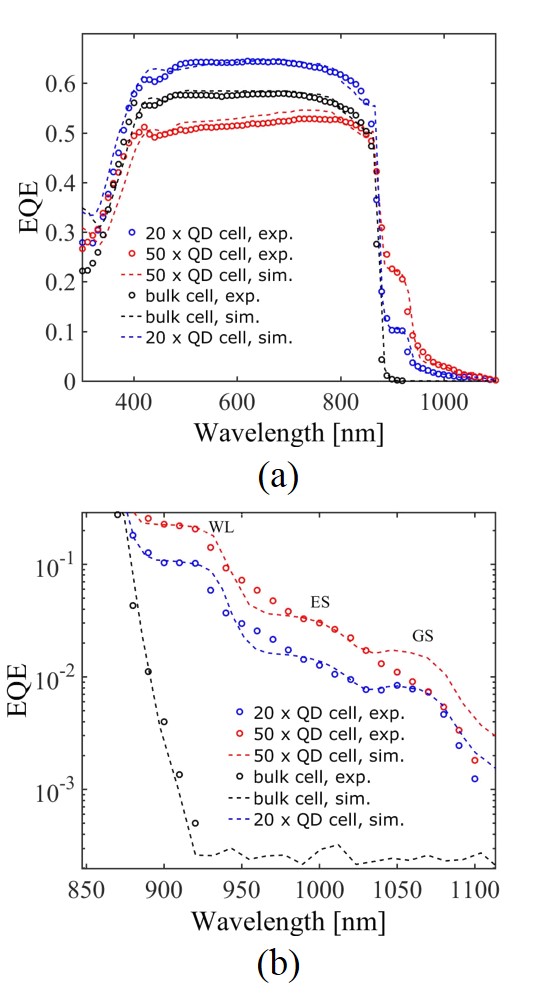}}
\caption{\label{Fig_EQEw} (a) Measured and simulated EQE spectra of the wafer-based samples. (b) Detail of the EQE spectra (in logarithmic scale) in the QD wavelength range.}
\end{figure}

Finally, the device-level study of the light-trapping enhanced QD cells is complemented by the design of photonic gratings for broadband antireflection and light diffraction which have been numerically simulated with the Rigorous Coupled-Wave Analysis (RCWA) method \cite{GD_Calc}.

\section{Results and Discussion}
\subsection{Wafer-based cells}
\label{sec_results_1}
Fig.\,\ref{Fig_EQEw} reports the measured EQE spectra of regular and QDSCs. The $20\times$ and $50\times$ QDSCs exhibit the expected long-wavelength EQE extension. Integration of the EQE over the AM1.5G sun spectrum for wavelengths above 880nm provides a current of 0.34 mA/cm$^2$ (i.e. 0.017 mA/cm$^2$ per QD layer) and 0.71 mA/cm$^2$ (i.e. 0.014 mA/cm$^2$ per QD layer) for the $20\times$ and $50\times$ QD layer cells, respectively. In the 400 nm - 800 nm wavelength range the $20\times$ QD cell has higher EQE than the one of the regular GaAs cell due to the higher collection efficiency provided by the intrinsic portion of the emitter which hosts the QD stack. In the $50\times$ QD cell, since the intrinsic region has more than doubled thickness that the one in the $20\times$ QD cell, an even higher EQE is expected in the same wavelength range. Instead, the measured EQE spectra of the $50\times$ QD cell shows a marked decrease in the 400 - 800 nm window with respect to the $20\times$ QD cell and to the regular GaAs cell. The short circuit current density estimated by integrating the measured EQE over the AM1.5G spectrum results as 17.58 mA/cm$^2$ for the regular cell, 19.95 mA/cm$^2$ for the $20\times$ QD cell, and 18.55 mA/cm$^2$ for the   $50\times$ QD cell. Finally, the measured $I-V$ characteristics (see figure S1 in the Supplementary Material) pinpointed a large and almost identical reduction of $V_\up{oc}$ (0.65 V) for both the QD solar cells with respect to the regular one (1.04 V). 

Such results can be understood with the help of the model described in Sec.\ref{model}, analysing the influence of the various recombination mechanisms in the different regions of the cell on the photovoltaic parameters. To this aim, non-radiative recombination is assumed to take place only from the extended states and is modeled according to Shockley Read Hall (SRH) theory. SRH recombination lifetimes were initially set in each region according to the doping type and level following the empirical model in \cite{2014Lumb_JAP}. For the adopted $n-p^+$ deep junction design, the most relevant parameters for the cell operation turn to be the hole SRH lifetime in the lighlty $n-$doped part of the emitter and the SRH lifetime in the intrinsic barrier layers of the QD stack, whose values have been estimated based on EQE and $I-V$ fitting \footnote{In this regard, it is worth noticing that different samples were processed from the same wafer for EQE and $I-V$ measurements. A slight discrepancy between the $J_\up{sc}$ extracted from the $I-V$ characteristics (see Supplementary Material) and the $J_\up{sc}$ estimated from the integration of the EQE spectrum was observed and attributed to some cross-wafer variability. Despite this, the fitting allowed to identify clear trends, as discussed in the text.}.

\begin{figure}
\centerline{\includegraphics[width=1\columnwidth]{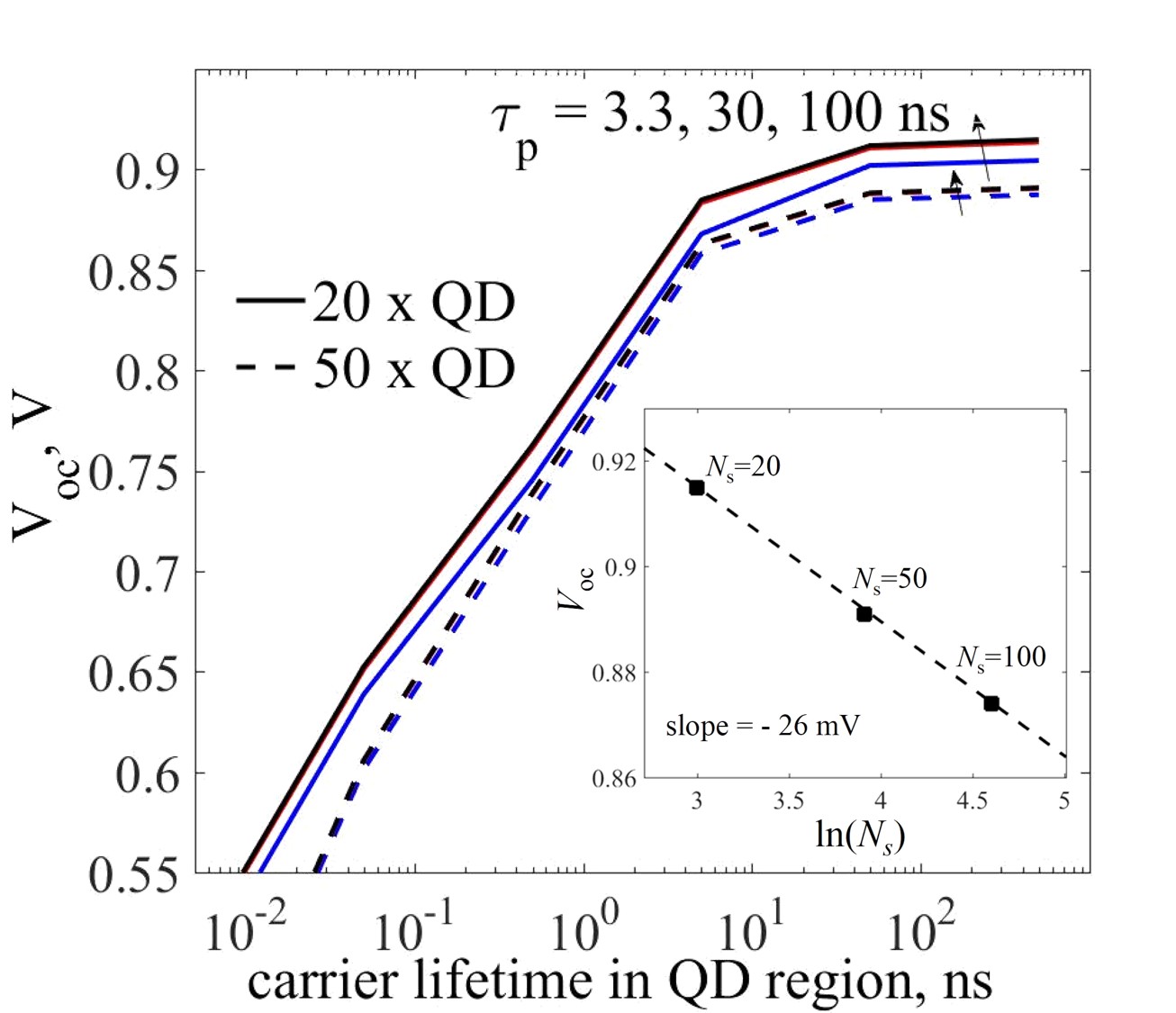}}
\caption{\label{Fig_Voc} Behavior of $V_\up{oc}$ for the $20\times$ (solid lines) and $50\times$ (dashed lines) QD cells as a function of the non radiative recombination lifetime in the interdot layers of the QD stack, for different values of hole SRH lifetime in the doped portion of the emitter ($\tau_p$). The insets shows the scaling of $V_\up{oc}$ with the natural logarithm of the number of QD layers ($N_\up{s})$: symbols are results of simulations, the dashed line is the linear fitting with 26 mV slope according to $V_\up{oc}=V_\up{T}\ln(J_\up{sc}/J_0)\propto -V_\up{T}\ln(N_\up{s})$.}
\end{figure}
\begin{figure}
\centerline{\includegraphics[width=1\columnwidth]{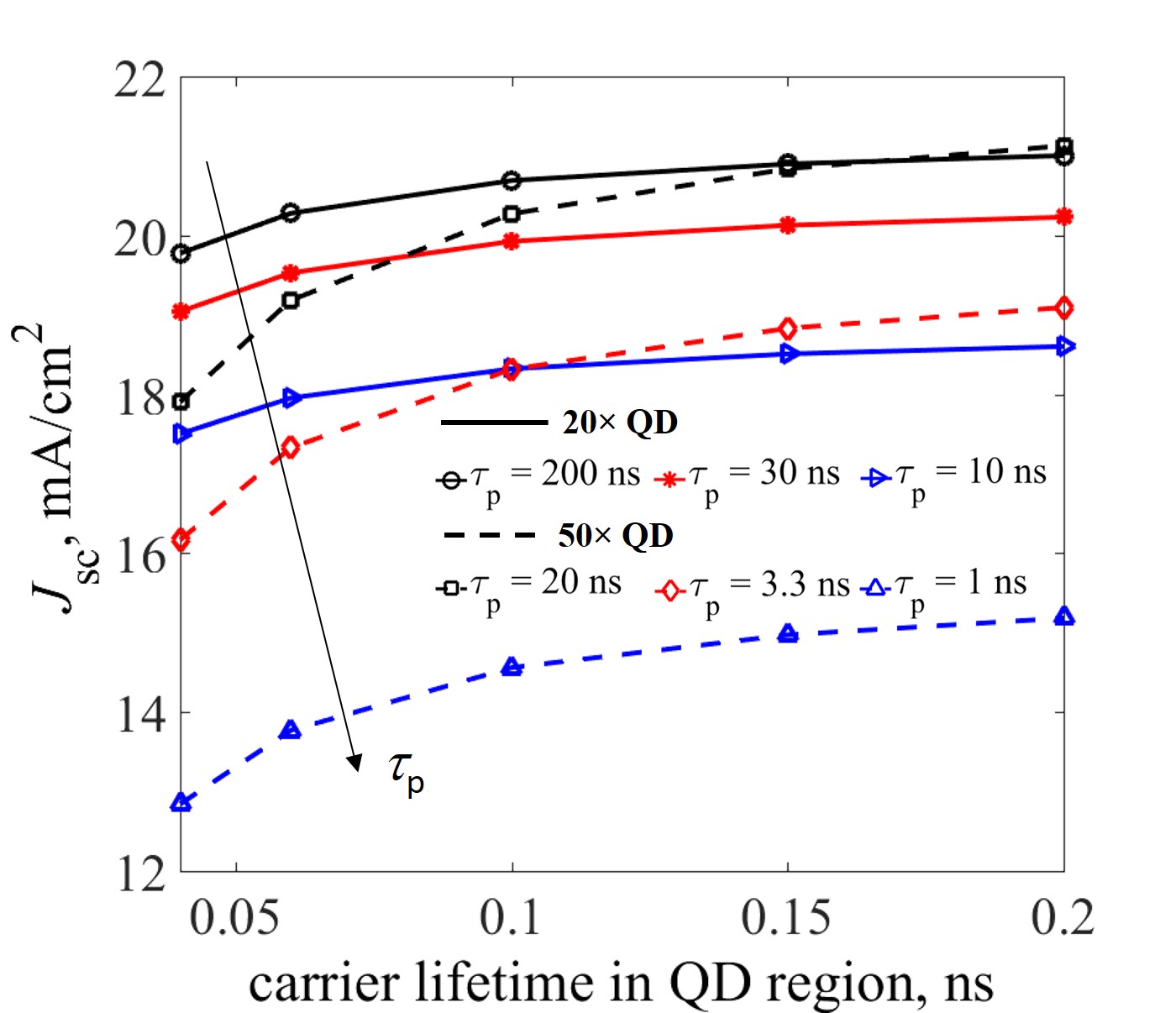}}
\caption{\label{Fig_Jsc} Behavior of $J_\up{sc}$ for the $20\times$ (solid lines) and $50\times$ (dashed lines) QD cells as a function of the non radiative recombination lifetime in the interdot layers of the QD stack, for different values of hole SRH lifetime in the doped portion of the emitter ($\tau_p$).}
\end{figure}

As shown in Fig.\,\ref{Fig_Voc}, the $V_\up{oc}$ is very sensitive to the SRH lifetime in the interdot layers, while it is only marginally affected by the hole SRH lifetime in the lightly doped region of the emitter. As long as the SRH lifetime is significantly larger than the QD radiative lifetime (1 ns), the $V_\up{oc}$ remains high with calculated limiting values (for negligible SRH recombination) of 915 mV ($20\times$ QD) and 891 mV ($50\times$ QD) against 1.065 V of the regular cell. This provides an indication of the inevitable $V_\up{oc}$ penalty compared to the single-junction GaAs cell induced by the inclusion of QDs, under the hypothesis of ideal material quality. The reason is that under forward bias, because of radiative recombination through the confined states, QDs act as trap centers whose activation energy increases as the level confinement energy decreases, i.e. as thermally-assisted escape becomes more effective \cite{2013Gioannini_JPV,2016Cappelluti_SEMSC}. In line with this interpretation, \cite{2012Jolley_PPRA} showed that the dark current of a InAs/GaAs QDSC has a lower thermal activation energy (about 100 mV) compared to the reference single-junction cell. When non radiative recombination is discarded, simulations predict a $V_\up{oc}$ offset with respect to the GS badgap of 0.255 V for the 20$\times$ QD cell and 0.279 V for the $50\times$ QD cell. The bandgap-voltage offset scales linearly with the natural logarithm of the number of layers in the QD stack ($N_\up{s}$) with slope equal to 26 mV, as shown in the inset in Fig.\,\ref{Fig_Voc}. Since $V_\up{oc}=V_\up{T}\ln(J_\up{sc}/J_0)$,  $J_0$ being the reverse saturation current and $V_\up{T}=26$ mV, and $J_\up{sc}$ is weakly affected by the number of QD stacks, this denotes an almost linear increase of the radiative recombination with the number of QD layers.

The estimated bandgap-voltage offset is in very good agreement with the value of 0.3 V reported in \cite{2012Tanabe_APL} for a collection of experimental data of high-quality QDSCs, considering that real devices likely have a slightly higher penalty due to some residual amount of non radiative recombination. In comparison, the calculated offset for the GaAs cell in the radiative limit is 0.36 V, while 0.4 V is the value of state-of-art wafer-based single-junction GaAs cells \cite{2012Tanabe_APL,2011King_PPRA} and 0.30 V that one of the highest efficiency thin-film GaAs cells. It is worth reminding that the bandgap-voltage offset is found to be relatively invariant with respect to the forbidden bandgap, both in experiments and theoretical calculations: in \cite{2011King_PPRA}, a maximum decrease of 50 mV for bangap ranging from 2.0 to 0.7 eV was predicted based on detailed balance calculations. 

The fairly linear dependence of $V_\up{oc}$ with the GS energy bandgap previously reported in \cite{2012Tanabe_APL,2013Gioannini_JPV} and the bandgap-voltage offset values calculated here (when radiative recombination is the only carrier loss mechanism) demonstrate that the $V_\up{oc}$ offset of 0.3 V with respect to the GS bandgap is a useful - albeit conservative - metric to gauge the QDSC material quality.
In fact, in the samples under study the observed reduction of $V_\up{oc}$ can be attributed to crystal quality degradation in the QD stack: as seen in Fig.\,\ref{Fig_Voc}, a SRH lifetime of 50 ps and 120 ps in the $20\times$ QD and $50\times$ QD cells provide a good fitting of the measured $V_\up{oc}$ of 0.65 V.  On the other hand, as illustrated in Fig.\,\ref{Fig_Jsc}, the $J_\up{sc}$ is extremely sensitive to the SRH hole lifetime in the thick, lightly doped $n$-type emitter. From the EQE and $I-V$ fitting of single-junction and $20\times$ QD cells, the SRH hole lifetime in the $n$-doped emitter is found to range between 15 ns and 200 ns (for the sake of reference, the radiative lifetime in the $n$-doped emitter is about 50 ns) denoting overall a reasonable, although improvable, crystal quality.  For the $50\times$ QD cell, the extracted SRH hole lifetime decreases to about 3 ns, suggesting a degradation of the emitter quality due to strain-induced defects caused by the stacking of a larger number of QD layers. Simulations reproduce the measured EQE (Fig.\,\ref{Fig_EQEw}) and $I-V$ characteristics (figure S1 in the Supplementary Material) with very good accuracy. The overall scope of results shows that the adopted physics-based model describes accurately and reliably (taking also into account the limited number of parameters used for the sake of fitting) the physics of QD solar cells, providing a useful means not only for analysis purpose but also to devise design guidelines, as done in Sec.\,\ref{sec_results_3} for the light-trapping enhanced thin-film cells.

\begin{figure}
\centerline{
\includegraphics[width=0.95\columnwidth]{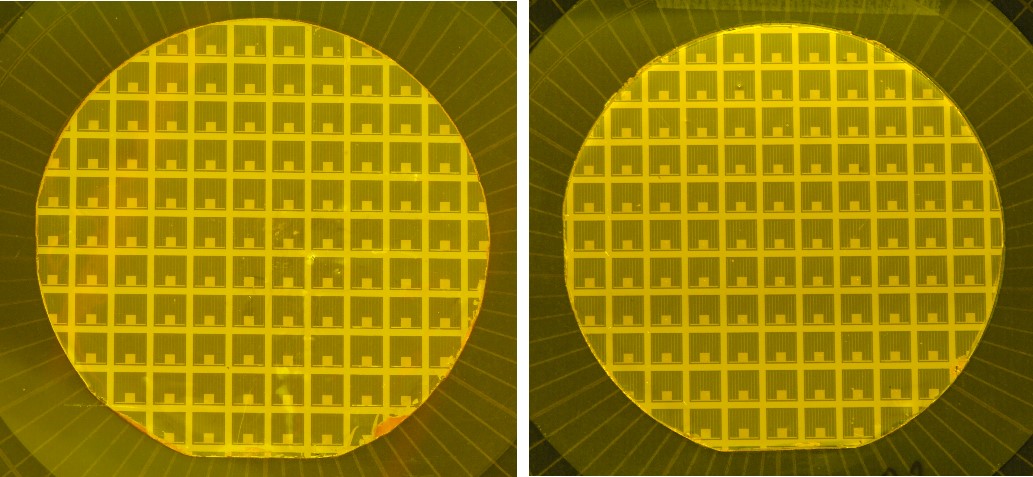}}
\caption{\label{Fig5} Single-junction GaAs (left) and $20\times$ QD InAs/GaAs (right) thin-film cells from epitaxial lift-off of 3-inch diameter epi-structures.}
\end{figure}

\subsection{ELO thin-film cells with planar reflector}
\label{sec_results_2}
\begin{figure}[ht]
\centerline{\includegraphics[width=.9\columnwidth]{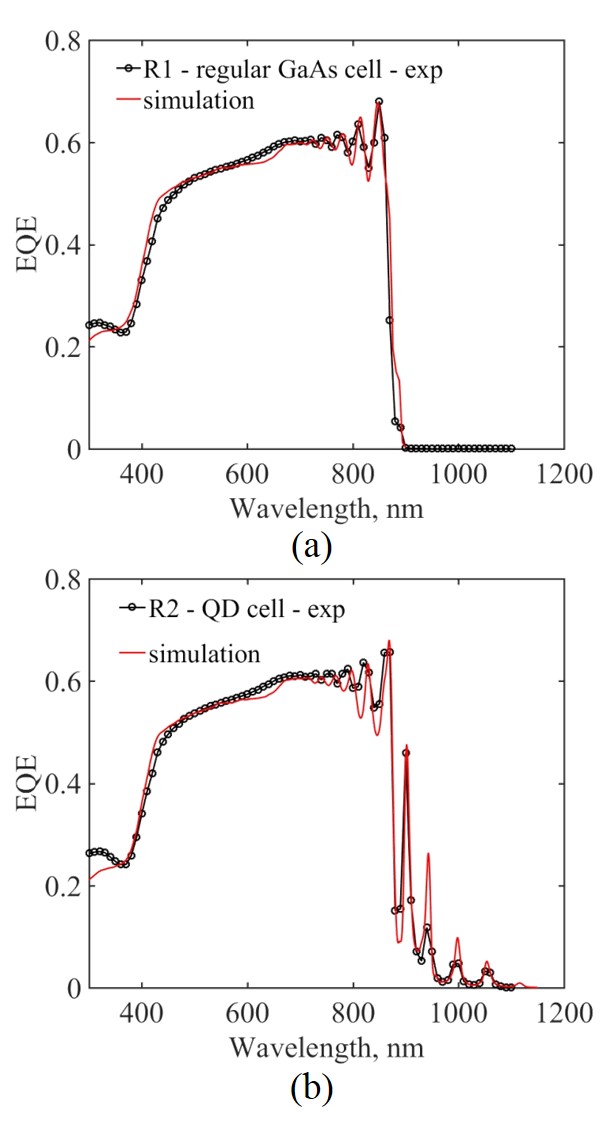}}
\caption{\label{Fig6} Measured and simulated EQE spectra of thin-film single-junction GaAs solar cells (top) and thin-film QD solar cells (bottom).}
\end{figure}

Full 3-inch films were released from their substrates by ELO and processed into cells (area=0.25 cm$^2$) with planar gold mirror on the backside, as shown in Fig.\,\ref{Fig5}. EQE spectra of the thin-film cells are reported in Fig.\,\ref{Fig6}. Clear resonant cavity effects are observed - both in experiments and simulations - in the low absorption wavelength range ($\lambda > 750$ nm) involving GaAs band-edge and QD interband transitions. Most important, the photogeneration in the QD wavelength range is nearly doubled in the thin-film configuration with respect to the wafer-based configuration: for $\lambda>880$ nm, the ELO QD cell provides a short circuit current density of about 0.62 mA/cm$^2$ against the 0.34 mA/cm$^2$ in the wafer-based QDSC (Fig.\,\ref{Fig_EQEw}). In the GaAs range (400 - 800 nm) the response of the ELO QD and regular GaAs cells is comparable, with a slightly higher EQE in the QD cell due to the intrinsic portion of the emitter, as already discussed for the wafer-based cells. From the EQE fitting, the SRH hole lifetime in the emitter is estimated around 40 ns for both the samples.

\begin{figure}[ht]
\centerline{\includegraphics[width=.9\columnwidth]{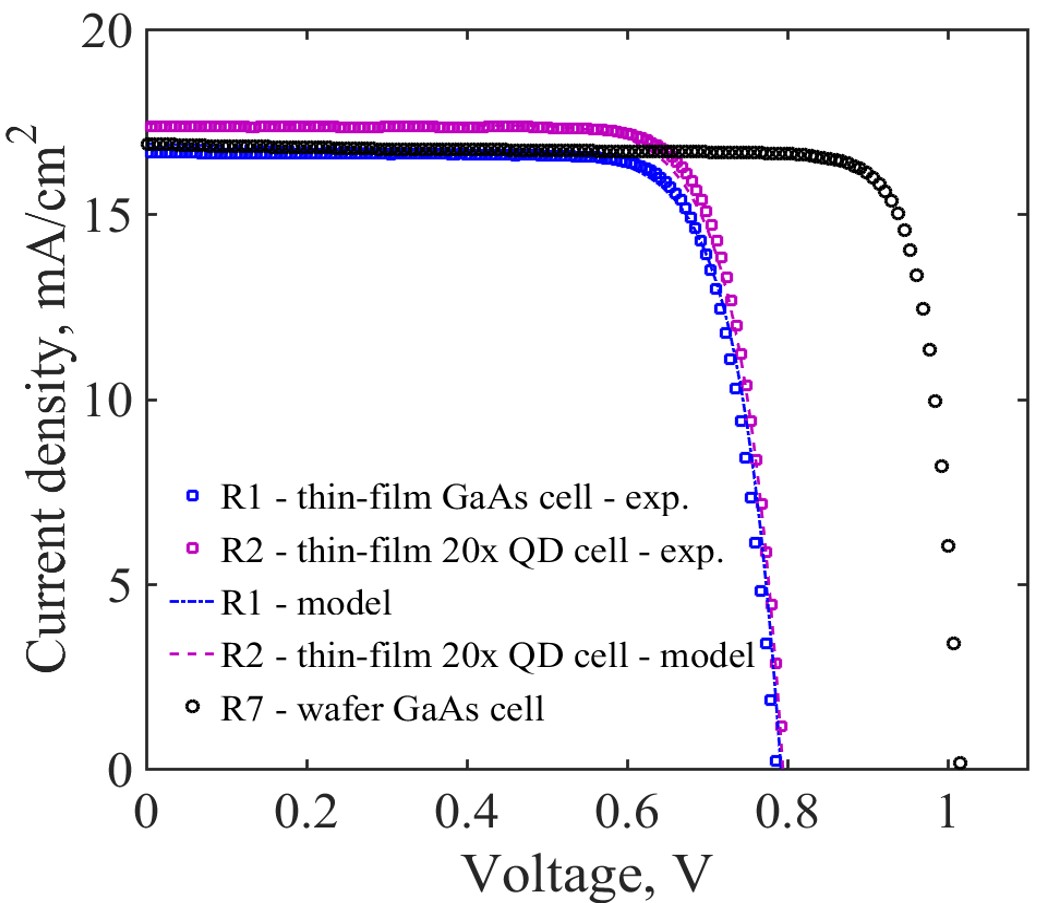}}
\caption{\label{Fig7}  Current-Voltage characteristics of wafer-based regular GaAs cell and thin-film GaAs and QD cells measured under AM1.5G illumination.}
\end{figure}

\begin{figure}
\centerline{\includegraphics[width=1\columnwidth]{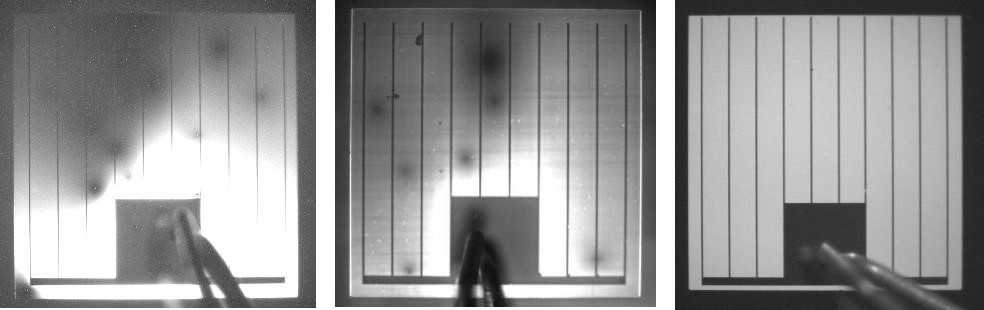}}
\caption{\label{Fig8} Representative electroluminescence images of the regular (left) and QD (middle) thin-film cells in comparison to that of a substrate based cell (right). The thin-film cells showing point defects and misfit dislocation lines typically require currents in the 40-80 mA range while the substrate based cells showing homogeneous luminescence require currents in the 0.5-5 mA range to obtain an image.}
\end{figure}

Fig.\,\ref{Fig7} shows the $I-V$ characteristics of the ELO cells and of the baseline wafer-based GaAs cell. The photovoltaic figures of the different cells are summarized in Table\,\ref{table}. The $J_\up{sc}$ of the ELO QDSC is increased by about 0.7 mA/cm$^2$ with respect to the regular GaAs ELO cell as expected from the QD enhanced contribution and slighlty improved above-gap collection observed in the EQE measurements. Furthermore, ELO and wafer-based GaAs cells have nearly identical $J_\up{sc}$, denoting that the ELO process did not have any detrimental effect on carrier collection. It is worth noticing that the thin-film QD cell shows an absolute improvement of efficiency of 0.6\% with respect to the thin-film single-junction counterpart. In fact, whereas the short circuit current of the QD cell is higher than that one of the regular thin-film cell, their $V_\up{oc}$ is comparable ($\approx 0.8$ V), with a significant reduction with respect to the regular wafer-based cell ($V_\up{oc}\approx 1$ V). The almost identical $V_\up{oc}$ of the regular and ELO QD cells makes not possible to draw any conclusion about the actual $V_\up{oc}$ penalty induced by QDs in this second generation of devices. We can only infer an upper bound of about 240 mV taking as reference the 1.04 V of the wafer-based single-junction cell.  This denotes (according to the simulated trends in Fig.\,\ref{Fig_Voc}) a significant improvement of the SRH lifetime ($\approx 1$ ns) in the interdot layers with respect to the previous generation of epilayers used for the wafer-based cells. The observed $V_\up{oc}$ degradation - totally unexpected for the regular GaAs cell - is attributed to lattice defects propagating from the substrate to the epilayers. The dark current of all the thin-film cells, regardless of the presence of QDs, is in fact dominated by the presence of a diode-like shunt defect, with reverse saturation current density of about 4 nA/cm$^2$ and ideality factor of 2, as shown from the $I-V$ fit in Fig.\,\ref{Fig7}. As further support of such conclusion, Fig.\,\ref{Fig8} shows the typical electroluminescence of the regular and QD thin-film cells in comparison to that one of a regular wafer-based cell. Both particle point-like defects and misfit dislocation lines can be clearly identified in the electroluminescence images of the thin-film ELO cells, demonstrating that the lower $V_\up{oc}$ and FF of the thin-film cells is related to random particle like defects and residual strain already present in the epi-structures prior to ELO processing. Further investigations demonstrated that such defect originated from a not sufficient quality of the initial substrates. These defects were extended during the more elaborative thin-film cell processing procedures. Using appropriate quality substrates yielding state-of-the-art epi-structures, ELO produced thin-film cells typically show equal performance or even outperform their wafer based counterparts \cite{2008Tatavarti_PVSC,2009Bauhuis_SEMSC,2011Kayes_PVSC} and in fact hold the current world record for a single junction GaAs cell \cite{2017Green_SolarTable}.

\begin{table}[h]
\renewcommand{\arraystretch}{1.2}
\caption{\label{table} Cell parameters of wafer-based and thin film ELO cells extracted from the $J-V$ characteristics in Fig.\ref{Fig7} }
\tabcolsep=0.12cm
\begin{tabular}{l|cccc}
Device & $J_\up{sc}$ (mA/cm$^2$) & $V_\up{oc}$ (V) & FF (\%) & $\eta$ (\%) \\
\hline
regular wafer-based & 16.9 & 1.017 & 84.7 & 14.6 \\
regular thin-film & 16.7 & 0.786 & 78.7 & 10.3 \\
QD thin-film & 17.4 & 0.796 & 78.7 & 10.9 \\
\hline
\end{tabular}
\end{table}

\subsection{Photonic management approaches through cell nanostructuring}
\label{sec_results_3}
ELO thin-film cells provide a suitable platform for the implementation of light-trapping structures aimed at maximizing the IR harvesting at the GaAs bandedge and QD wavelengths. We use here the device-level model introduced in Sec.\,\ref{model} to quantify the impact of the QD photogeneration enhancement on the performance of thin-film QD cells integrating photonic structures for light management.

We take as reference the epilayer structure exploited for the 50 $\times$ QD solar cell analysed in Sec.\ref{sec_results_1}, with 20, 50, or 100 QD layers uniformly distributed in the $\approx 1\,\mu$m thick undoped part of the emitter. The analysis is done by assuming negligible non radiative recombination and surface recombination. All the other material parameters are those extracted from the analysis of the fabricated devices. Under the assumption of defect-free material, photon recycling becomes a relevant effect with significant influence on the open circuit voltage and the ultimate  efficiency \cite{2012Miller_JPV,2013Wang_JPV}. This is well known for single-junction GaAs cells and may be expected to have a similar role in QD solar cells. In \cite{2017Cappelluti_ESPC} we have shown through physics-based simulations that under the hypothesis of very efficient photon recycling, QDSCs might reach efficiency higher than 30\%. However, from a practical standpoint, whereas in a single-junction GaAs cell a planar reflector is sufficient to achieve strong photon recycling (since photons are re-emitted isotropically), the QD cell needs at the same time texturing of the solar cell surface(s) for light-trapping. Surface recombination at the semiconductor textured interfaces may become critical, because very low surface recombination velocity is needed to fully exploit the photon recycling potential. Thus, we take as conservative assumption to neglect the photon-recycling effect, somewhat compensating for the fact that we are also neglecting surface recombination and we focus only on the analysis of the impact of light trapping. Under these assumptions the predicted $V_\up{oc}$ is 0.925 V, 0.902 V, 0.884 V, for the $20\times$, $50\times$, $100\times$ wafer-based QD cells, respectively. The voltage offset from the GS energy gap is reduced of about 10 mV compared to the wafer-based configuration and scales with the number of QD layers according to the same $-V_\up{T}\ln{N_\up{s}}$ law.

\begin{figure}
\centerline{\includegraphics[width=.95\columnwidth]{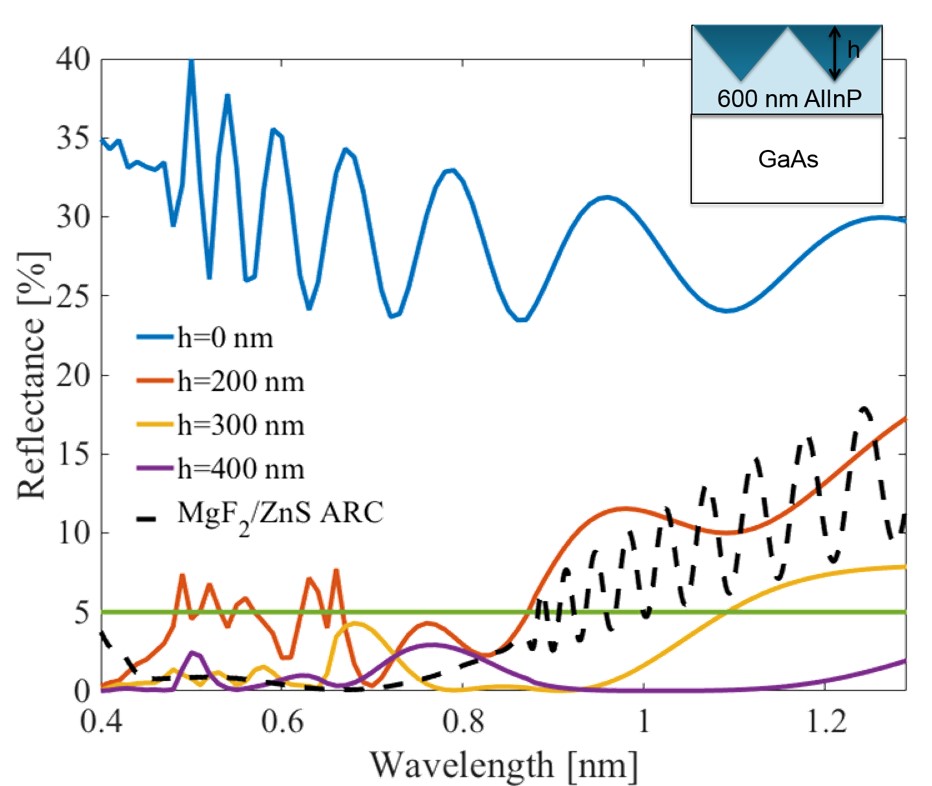}}
\caption{\label{Fig_ARC} Optimization of AlInP based pyramidal ARC. For the sake of comparison the reflectivity of a standard double-layer ARC is shown.}
\end{figure}

Enhancement of the QD photocurrent requires effective photonic management to maximize photon trapping and to minimize  photon reflection loss in the QD wavelength range. In this respect, conventional two-layer ARCs usually exploited for triple-junction cells are not sufficient, since the power reflection requirements at long wavelengths are quite relaxed owing to the over-current brought by the bottom cell. To achieve very low reflectance across the QD range, moth-eye broadband ARCs can be realized by nano-patterning the cell front-surface \cite{2010Tommila_SEMSC}. As an example, Fig.\,\ref{Fig_ARC} shows the RCWA-calculated reflectivity for pyramids nanoimprinted on a 600 nm thick AlInP window layer of a cell with geometry similar to that under study. The pyramidal grating has a period of 200 nm, and with height of about 400 nm provides a reflectivity
which is comparable in average to a conventional MgF$_2$/ZnS ARC in the GaAs band gap range, and extremely low - with values well below 1\% - over the QD wavelength range. Based on this, we assume in the following that the power loss due to the cumulative effect of reflection and shadowing loss is limited to 5\%.

\begin{figure}
\centerline{\includegraphics[width=1\columnwidth]{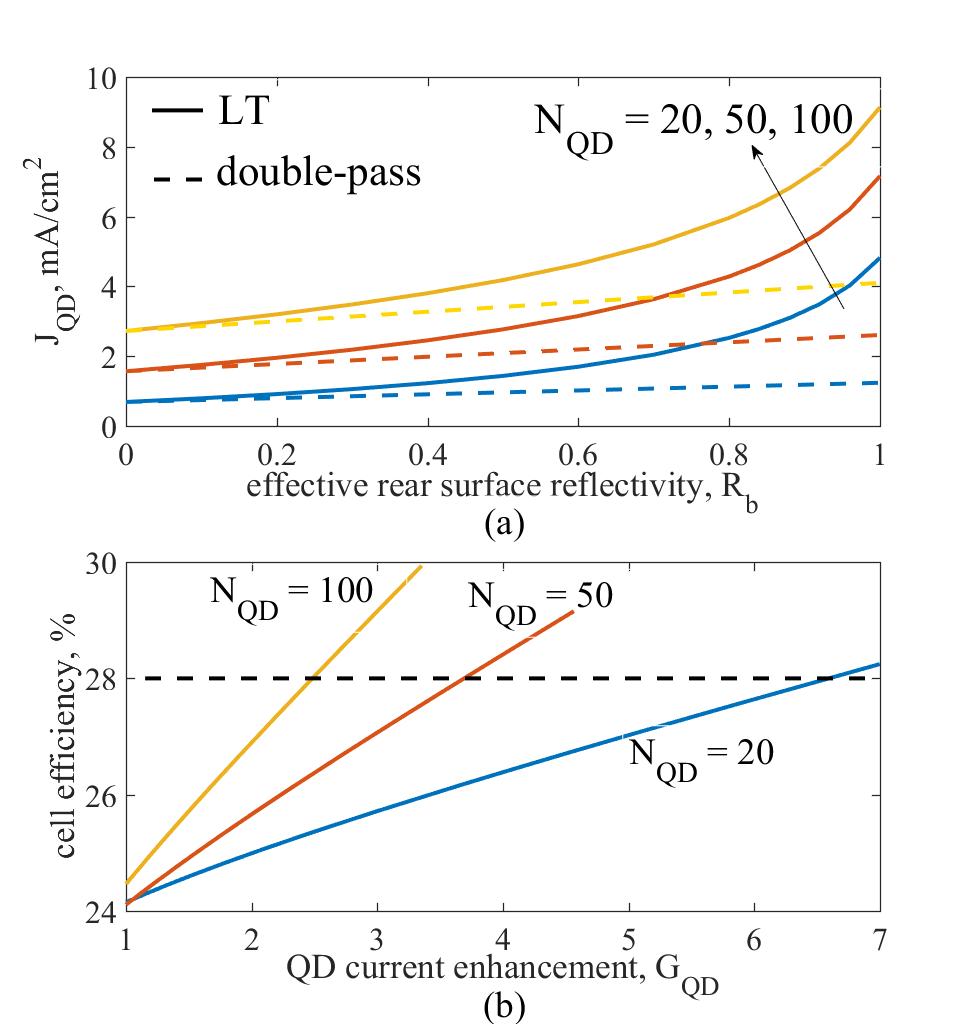}}
\caption{\label{Fig_Eff} (a) QD current density as a function of the effective rear surface reflectivity for a thin-film cells with planar mirror - double-pass configuration – and  thin-film cell with textured mirror - light-trapping configuration - using different numbers of QD layers. (b) Predicted cell efficiency as a function of the QD current enhancement with respect to the wafer-based configuration. The $R_\up{b}=0$ point identifes the wafer-based configuration.}
\end{figure}

Fig.\ref{Fig_Eff}(a) analyses the fraction of short-circuit current provided by the QDs ($J_\up{QD}$) for cells using increasing number of QD layers in the planar thin-film configuration (double-pass) and in the light-trapping configuration (LT), according to the incoherent multiple reflection model in Sec.\ref{model}. $J_\up{QD}$ is estimated by integrating the EQE in the QD wavelength range ($\lambda>880$ nm) over the AM1.5G solar spectrum. The LT scheme provides a remarkable increase of the QD current contribution compared to the wafer-based configuration ($R_\up{b}=0$) and to the planar (double-pass) one, The increase obviously tends to saturate as the number of QD layers increases. By defining the QD current enhancement, $G_\up{QD}$ as the ratio between the collected current $J_{\up{QD}}$ at a certain value of $R_\up{b}$ and $J_{\up{QD}}$ measured in the wafer-based configuration ($R_\up{b}=0$), we observe that in the ideal condition of $R_\up{b}=1$, the maximum achievable gain is about 7, 4.5, and 3.4 for the $20\times$, $50\times$, $100\times$ QD cells, respectively. The impact in terms of power conversion efficiency is shown in Fig.\,\ref{Fig_Eff}(b). An enhancement of about 4\% absolute efficiency is attained by the thin-film $20\times$ QD cell with respect to the wafer-based configuration provided that a sevenfold increase of the QD current can be reached, yielding the $20\times$ QD cell efficiency beyond 28\%. As previously mentioned, a further improvement of 2\%-3\% may be expected if photon recycling can be fully exploited. The impact of light-trapping becomes even stronger as the number of layers in the QD stack increases,  provided - obviously - that the material quality can be preserved. In contrast, it is worth noticing that in the conventional wafer-based configuration (corresponding to $G_\up{QD}=1$ in Fig.\,\ref{Fig_Eff}(b)) increasing the number of QD layers - within realistic ranges - does not provide any substantial advantage in terms of attainable efficiency. Similar conclusions were also drawn in \cite{2014Mellor_SEMSC} for intermediate band QDSCs working in the IB operating regime on the basis of detailed balance calculations.

\begin{figure}
\centerline{\includegraphics[width=.95\columnwidth]{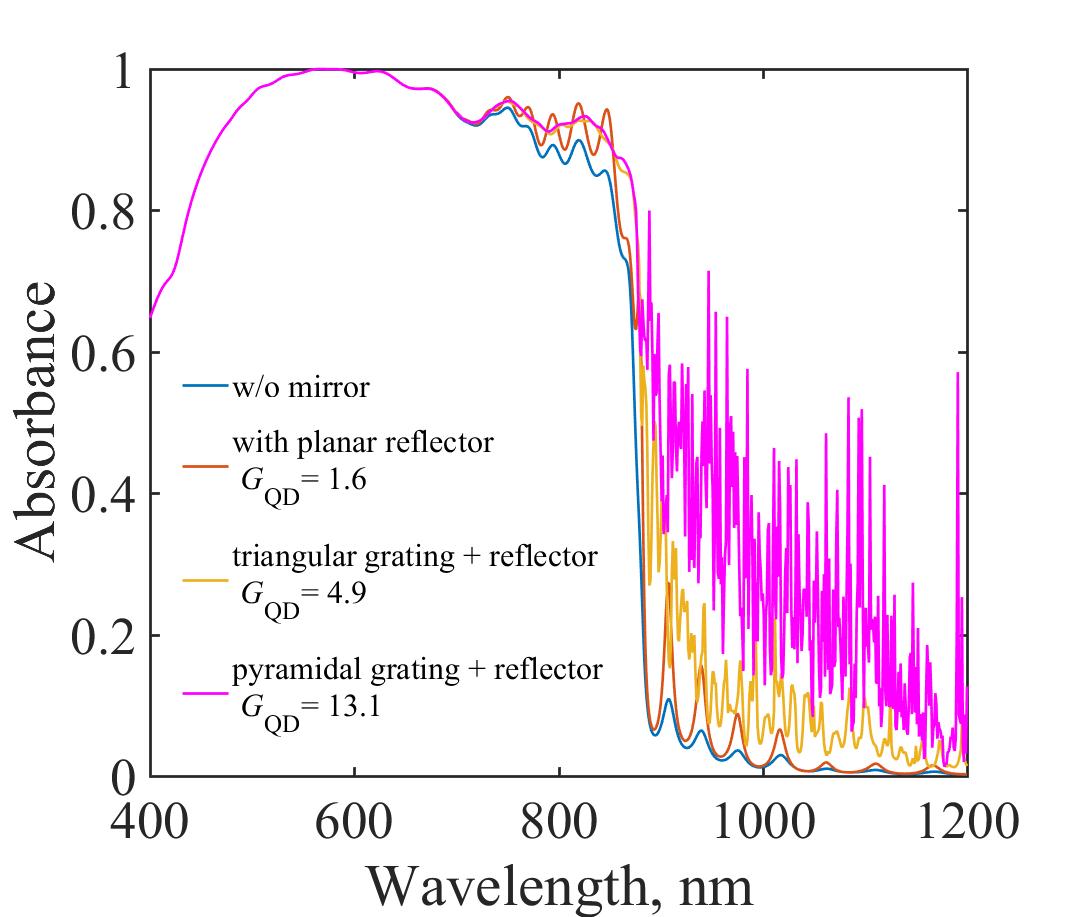}}
\caption{\label{Fig_Diffr} Calculated absorbance spectra for different photonic configurations.}
\end{figure}

In our technology roadmap, efficient light-trapping shall be implemented by patterning a diffraction grating on the rear surface of the cell. Differently from ARCs, light diffraction requires a grating period larger than the incident wavelength. The micro-structured grating excites high order diffraction modes propagating outside of the cell escape cone and thus increases the optical path length at low absorbing wavelengths. Several designs based on lamellar gratings with triangular cross-section and bi-periodic pyramidal gratings imprinted on the widegap (AlInP) back surface field layer were studied in \cite{2016Musu_OSA,2017Cappelluti_ESPC} with RCWA simulations and led to identify optimized geometries with grating period around 2$-$3 $\mu$m and aspect ratio (height/period) on the order of 0.3$-$0.4. An example of calculated absorbance spectra for a $20\times$ QD cell exploiting different photonic configurations is shown in Fig.\,\ref{Fig_Diffr}. With the micro-structured pyramidal grating, an enhancement of QD photocurrent ($G_\up{QD}$) of about 13 times with respect to a thin-film cell without mirror is predicted, which is largely compatible with the QD photocurrent enhancement of 7 identified in Fig.\,\ref{Fig_Eff}(b) to surpass 28\% efficiency with 20 QD layers. Simulations presented in Fig.\,\ref{Fig_Diffr} were done under the hypothesis of a lossless mirror conformally deposited over the micro-structured grating.  Further studies (not reported here) show that a similar enhancement can be achieved when realistic metal loss are taken into account, by including between the micro-structured grating and the metal a low index planarizing dielectric layer that shall also passivate the semiconductor surface.

\section{Conclusions}
We have reported thin-film InAs/GaAs QD solar cells fabricated by epitaxial lift-off of 3-inch wafers containing
QD epi-structures with high ($\approx 8\times 10^{10}$ cm$^{-2}$) in-plane QD density. A clear increase of EQE in the long wavelength region is observed in ELO  QD cells with planar gold reflector, yielding  doubled QD current contribution at short circuit. Moreover, ELO QD cells show nearly identical collection efficiency compared to the baseline ELO GaAs cell, confirming ELO as very well suited to epistructures containing QD layers. Unfortunately, the $V_\up{oc}$ of about 0.8 V of the present generation of ELO cells - either single-junction and QD cells -  was impaired by defects originated from a not sufficient quality of the initial substrates.

An intensive simulation study has also been presented with the twofold objective of quantifying the impact of fundamental and non-fundamental (i.e. technologically related) processes on the cell photovoltaic performance and investigating the feasibility of high efficiency QDSCs operating in the thermally-limited regime. To this aim, we have adopted a classical drift-diffusion model coupled with a phenomenological rate equations model to account for QD kinetics \cite{2013Gioannini_JPV}. The comparison between measurements and simulations demonstrate that such modelling approach describes very accurately the cell physics, providing a useful means for design purpose. For the QDs under study we predict that $V_\up{oc}$ higher than 0.9 V is achievable if non radiative recombination is substantially suppressed. As corollary - yet important - result of this analysis, we provide a theoretical background to the empirical finding in \cite{2012Tanabe_APL} of a $V_\up{oc}$ offset of 0.3 V from the ground state energy gap in QD cells. Thus, at least with the aim at demonstrating high-efficiency InAs/GaAs QD cells under unconcentrated light, ambient temperature, shallow QDs shall be preferred to minimize the $V_\up{oc}$ penalty.

Finally, with the support of the validated electro-optical model and of rigorous electromagnetic simulations, we have investigated the efficiency potential and practical development of thin-film InAs/GaAs QD cells with photon gratings. It is shown that by proper light management, through the use of broadband antireflection and diffraction gratings, the $20\times$ QD cell may achieve about sevenfold increase of the QD near-infrared current and efficiency higher than 28\%. These cells can be fabricated using cost-effective and scalable fabrication processes such as epitaxial lift-off -as demonstraed in this work - for the thin-film processing, and nanoimprint lithography to pattern even subwavelength period gratings over large areas \cite{2010Tommila_SEMSC}.

The study has also evidenced that increasing the number of QD layers is an effective means to achieve high efficiency QDSCs only if combined with light-trapping. Photon-recycling - not included in this study - is expected to provide a further increase of the predicted efficiency. We believe that the promising experimental results and the guidelines provided by the numerical simulations presented in this work are important in view of demonstrating high efficiency InAs/GaAs QDSCs in the short term and we anticipate that they will be also useful in the longer term for intermediate band QDSCs.

\section{Acknowledgements}
The research was supported by the European Union's Horizon 2020 research and innovation program, under Grant Agreement 687253 – TFQD. http://tfqd.eu

\section*{References}
\bibliography{}

\begin{thebibliography}{10}
\expandafter\ifx\csname url\endcsname\relax
  \def\url#1{\texttt{#1}}\fi
\expandafter\ifx\csname urlprefix\endcsname\relax\def\urlprefix{URL }\fi
\expandafter\ifx\csname href\endcsname\relax
  \def\href#1#2{#2} \def\path#1{#1}\fi

\bibitem{2001Aroutiounian_JAP}
V.~Aroutiounian, S.~Petrosyan, A.~Khachatryan, K.~Touryan, Quantum dot solar
  cells, JAP 89~(4) (2001) 2268--2271.

\bibitem{2014Kerestes_PPRA}
C.~Kerestes, S.~Polly, D.~Forbes, C.~Bailey, A.~Podell, J.~Spann, P.~Patel,
  B.~Richards, P.~Sharps, S.~Hubbard, Fabrication and analysis of multijunction
  solar cells with a quantum dot (in) gaas junction, Progress in Photovoltaics:
  Research and Applications 22~(11) (2014) 1172--1179.

\bibitem{2016Ho_PPRA}
W.-J. Ho, Y.-Y. Lee, G.-C. Yang, C.-M. Chang, Optical and electrical
  characteristics of high-efficiency ingap/ingaas/ge triple-junction solar cell
  incorporated with ingaas/gaas qd layers in the middle cell, Progress in
  Photovoltaics: Research and Applications 24~(4) (2016) 551--559.

\bibitem{1997Luque_PRL}
A.~Luque, A.~Mart\'{\i}, Increasing the efficiency of ideal solar cells by
  photon induced transitions at intermediate levels, Phys. Rev. Lett. 78~(26)
  (1997) 5014--5017.

\bibitem{2006Green_Third}
M.~A. Green, et~al., Third generation photovoltaics, Springer, 2006.

\bibitem{2001Luque_TED}
A.~Luque, A.~Mart{\'\i}, L.~Cuadra, Thermodynamic consistency of sub-bandgap
  absorbing solar cell proposals, IEEE Transactions on Electron Devices 48~(9)
  (2001) 2118--2124.

\bibitem{2007Wei_APL}
G.~Wei, K.-T. Shiu, N.~C. Giebink, S.~R. Forrest, Thermodynamic limits of
  quantum photovoltaic cell efficiency, Applied Physics Letters 91~(22) (2007)
  223507.
\newblock \href {http://dx.doi.org/10.1063/1.2817753}
  {\path{doi:10.1063/1.2817753}}.

\bibitem{2012Tanabe_APL}
K.~Tanabe, D.~Guimard, D.~Bordel, Y.~Arakawa, High-efficiency inas/gaas quantum
  dot solar cells by metalorganic chemical vapor deposition, Appl. Phys. Lett.
  100 (2012) 1293905--193905--3.

\bibitem{2011Kayes_PVSC}
B.~M. Kayes, H.~Nie, R.~Twist, S.~G. Spruytte, F.~Reinhardt, I.~C. Kizilyalli,
  G.~Higashi, 27.6\% conversion efficiency, a new record for single-junction
  solar cells under 1 sun illumination, in: Photovoltaic Specialists Conference
  (PVSC), 2011 37th IEEE, IEEE, 2011, pp. 000004--000008.

\bibitem{2017Green_SolarTable}
M.~A. Green, Y.~Hishikawa, W.~Warta, E.~D. Dunlop, D.~H. Levi, J.~Hohl-Ebinger,
  A.~Ho-Baillie, Solar cell efficiency tables (version 50), Progress in
  Photovoltaics: Research and Applications 25~(7) (2017) 668--676.

\bibitem{2012Sakamoto_JAP}
K.~Sakamoto, Y.~Kondo, K.~Uchida, K.~Yamaguchi, Quantum-dot density dependence
  of power conversion efficiency of intermediate-band solar cells, Journal of
  Applied Physics 112~(12) (2012) 124515.

\bibitem{2014Mellor_SEMSC}
A.~Mellor, A.~Luque, I.~Tob{\'\i}as, A.~Mart{\'\i}, The feasibility of
  high-efficiency inas/gaas quantum dot intermediate band solar cells, Solar
  Energy Materials and Solar Cells 130 (2014) 225--233.

\bibitem{2015Okada_APR}
Y.~Okada, N.~Ekins-Daukes, T.~Kita, R.~Tamaki, M.~Yoshida, A.~Pusch, O.~Hess,
  C.~Phillips, D.~Farrell, K.~Yoshida, et~al., Intermediate band solar cells:
  Recent progress and future directions, Applied physics reviews 2~(2) (2015)
  021302.

\bibitem{2010Zhou_APL}
D.~Zhou, G.~Sharma, S.~Thomassen, T.~Reenaas, B.~Fimland, Optimization towards
  high density quantum dots for intermediate band solar cells grown by
  molecular beam epitaxy, Applied Physics Letters 96~(6) (2010) 061913.

\bibitem{2011Fujita_PVSC}
H.~Fujita, K.~Yamamoto, J.~Ohta, Y.~Eguchi, K.~Yamaguchi, In-plane quantum-dot
  superlattices of inas on gaassb/gaas (001) for intermediate band solar-cells,
  in: Photovoltaic Specialists Conference (PVSC), 2011 37th IEEE, IEEE, 2011,
  pp. 002612--002614.

\bibitem{2013Tutu_APL2}
F.~Tutu, J.~Wu, P.~Lam, M.~Tang, N.~Miyashita, Y.~Okada, J.~Wilson, R.~Allison,
  H.~Liu, Antimony mediated growth of high-density inas quantum dots for
  photovoltaic cells, Applied Physics Letters 103~(4) (2013) 043901.

\bibitem{2016Sameshima_APE}
K.~Sameshima, T.~Sano, K.~Yamaguchi, Self-formation of ultrahigh-density (1012
  cm-2) inas quantum dots on inassb/gaas (001) and their photoluminescence
  properties, Applied Physics Express 9~(7) (2016) 075501.

\bibitem{2011Akahane_PPSa}
K.~Akahane, N.~Yamamoto, T.~Kawanishi,
  \href{http://dx.doi.org/10.1002/pssa.201000432}{Fabrication of
  ultra-high-density inas quantum dots using the strain-compensation
  technique}, physica status solidi (a) 208~(2) (2011) 425--428.
\newblock \href {http://dx.doi.org/10.1002/pssa.201000432}
  {\path{doi:10.1002/pssa.201000432}}.
\newline\urlprefix\url{http://dx.doi.org/10.1002/pssa.201000432}

\bibitem{2012Sugaya_EES}
T.~Sugaya, O.~Numakami, R.~Oshima, S.~Furue, H.~Komaki, T.~Amano, K.~Matsubara,
  Y.~Okano, S.~Niki, Ultra-high stacks of ingaas/gaas quantum dots for high
  efficiency solar cells, Energy \& Environmental Science 5~(3) (2012)
  6233--6237.

\bibitem{2010Guimard_APL}
D.~Guimard, R.~Morihara, D.~Bordel, K.~Tanabe, Y.~Wakayama, M.~Nishioka,
  Y.~Arakawa, Fabrication of inas/gaas quantum dot solar cells with enhanced
  photocurrent and without degradation of open circuit voltage, Appl. Phys.
  Lett. 96~(20) (2010) 203507.

\bibitem{2016Cappelluti_SEMSC}
F.~Cappelluti, M.~Gioannini, A.~Khalili, Impact of doping on inas/gaas
  quantum-dot solar cells: a numerical study on photovoltaic and
  photoluminescence behavior, Solar Energy Materials and Solar Cells to
  appear~(-) (2016) --.

\bibitem{2010Jolley_APL}
G.~Jolley, H.~Lu, H.~Tan, C.~Jagadish, Electron-hole recombination properties
  of in$_{0.5}$ga$_{0.5}$as/gaas quantum dot solar cells and the influence on
  the open circuit voltage, Appl. Phys. Lett. 97 (2010) 123505--123505--3.

\bibitem{2013Gioannini_JPV}
M.~Gioannini, A.~Cedola, N.~Di~Santo, F.~Bertazzi, F.~Cappelluti, Simulation of
  quantum dot solar cells including carrier intersubband dynamics and
  transport, IEEE J. Photovoltaics 3~(4) (2013) 1271--1278.
\newblock \href {http://dx.doi.org/10.1109/JPHOTOV.2013.2270345}
  {\path{doi:10.1109/JPHOTOV.2013.2270345}}.

\bibitem{2011Bailey_APL}
C.~G. Bailey, D.~V. Forbes, R.~P. Raffaelle, S.~M. Hubbard, Near 1 v open
  circuit voltage inas/gaas quantum dot solar cells, Applied Physics Letters
  98~(16) (2011) 163105.

\bibitem{2012Bailey_JPV}
C.~G. Bailey, D.~V. Forbes, S.~J. Polly, Z.~S. Bittner, Y.~Dai, C.~Mackos,
  R.~P. Raffaelle, S.~M. Hubbard, Open-circuit voltage improvement of inas-gaas
  quantum-dot solar cells using reduced inas coverage, IEEE J. Photovoltaics
  2~(3) (2012) 269--275.

\bibitem{2016Smith_PVSC}
B.~L. Smith, M.~A. Slocum, Z.~S. Bittner, Y.~Dai, G.~T. Nelson, S.~D.
  Hellstroem, R.~Tatavarti, S.~M. Hubbard, Inverted growth evaluation for
  epitaxial lift off (elo) quantum dot solar cell and enhanced absorption by
  back surface texturing, in: Photovoltaic Specialists Conference (PVSC), 2016
  IEEE 43rd, IEEE, 2016, pp. 1276--1281.

\bibitem{2015Welser_book}
R.~E. Welser, A.~K. Sood, R.~B. Laghumavarapu, D.~L. Huffaker, D.~M. Wilt,
  N.~K. Dhar, K.~A. Sablon, The physics of high-efficiency thin-film iii-v
  solar cells, in: Solar Cells-New Approaches and Reviews, InTech, 2015, pp.
  247--276.

\bibitem{2016Cappelluti_PVSC}
F.~Cappelluti, M.~Gioannini, G.~Ghione, A.~Khalili, Numerical study of
  thin-film quantum-dot solar cells combining selective doping and
  light-trapping approaches, in: Photovoltaic Specialists Conference (PVSC),
  2016 IEEE 43rd, IEEE, 2016, pp. 1282--1286.

\bibitem{2008Tatavarti_PVSC}
R.~Tatavarti, G.~Hillier, A.~Dzankovic, G.~Martin, F.~Tuminello,
  R.~Navaratnarajah, G.~Du, D.~Vu, N.~Pan, Lightweight, low cost gaas solar
  cells on 4 ″epitaxial liftoff (elo) wafers, in: Photovoltaic Specialists
  Conference, 2008. PVSC'08. 33rd IEEE, IEEE, 2008, pp. 1--4.

\bibitem{2009Bauhuis_SEMSC}
G.~Bauhuis, P.~Mulder, E.~Haverkamp, J.~Huijben, J.~Schermer, 26.1\% thin-film
  gaas solar cell using epitaxial lift-off, Solar Energy Materials and Solar
  Cells 93~(9) (2009) 1488--1491.

\bibitem{1991Lush_SC}
G.~Lush, M.~Lundstrom, Thin film approaches for high-efficiency iii--v cells,
  Solar cells 30~(1-4) (1991) 337--344.

\bibitem{2012Miller_JPV}
O.~D. Miller, E.~Yablonovitch, S.~R. Kurtz, Strong internal and external
  luminescence as solar cells approach the shockley--queisser limit, IEEE
  Journal of Photovoltaics 2~(3) (2012) 303--311.

\bibitem{2013Wang_JPV}
X.~Wang, M.~R. Khan, J.~L. Gray, M.~A. Alam, M.~S. Lundstrom, Design of gaas
  solar cells operating close to the shockley--queisser limit, IEEE Journal of
  Photovoltaics 3~(2) (2013) 737--744.

\bibitem{2013Bennett_APL}
M.~F. Bennett, Z.~S. Bittner, D.~V. Forbes, S.~Rao~Tatavarti,
  S.~Phillip~Ahrenkiel, A.~Wibowo, N.~Pan, K.~Chern, S.~M. Hubbard, Epitaxial
  lift-off of quantum dot enhanced gaas single junction solar cells, Applied
  Physics Letters 103~(21) (2013) 213902.

\bibitem{2014Sogabe2014_APL}
T.~Sogabe, Y.~Shoji, P.~Mulder, J.~Schermer, E.~Tamayo, Y.~Okada, Enhancement
  of current collection in epitaxial lift-off inas/gaas quantum dot thin film
  solar cell and concentrated photovoltaic study, Applied Physics Letters
  105~(11) (2014) 113904.

\bibitem{2015Inoue_JPV}
T.~Inoue, K.~Watanabe, K.~Toprasertpong, H.~Fujii, M.~Sugiyama, Y.~Nakano,
  Enhanced light trapping in multiple quantum wells by thin-film structure and
  backside grooves with dielectric interface, IEEE Journal of Photovoltaics
  5~(2) (2015) 697--703.

\bibitem{2006Schermer_TSF}
J.~Schermer, G.~Bauhuis, P.~Mulder, E.~Haverkamp, J.~Van~Deelen,
  A.~Van~Niftrik, P.~Larsen, Photon confinement in high-efficiency, thin-film
  iii--v solar cells obtained by epitaxial lift-off, Thin Solid Films 511
  (2006) 645--653.

\bibitem{2014Gioannini_IET}
M.~Gioannini, A.~P. Cedola, F.~Cappelluti, Impact of carrier dynamics on the
  photovoltaic performance of quantum dot solar cells, Optoelectronics, IET
  9~(2) (2015) 69--74.
\newblock \href {http://dx.doi.org/10.1049/iet-opt.2014.0080}
  {\path{doi:10.1049/iet-opt.2014.0080}}.

\bibitem{2016Cappelluti_IET}
F.~Cappelluti, A.~Khalili, M.~Gioannini, Open circuit voltage recovery in
  quantum dot solar cells: a numerical study on the impact of wetting layer and
  doping, IET Optoelectronics 11~(2) (2016) 44--48.

\bibitem{2000Williamson_PRB}
A.~J. Williamson, L.~W. Wang, A.~Zunger,
  \href{https://link.aps.org/doi/10.1103/PhysRevB.62.12963}{Theoretical
  interpretation of the experimental electronic structure of lens-shaped
  self-assembled inas/gaas quantum dots}, Phys. Rev. B 62 (2000) 12963--12977.
\newblock \href {http://dx.doi.org/10.1103/PhysRevB.62.12963}
  {\path{doi:10.1103/PhysRevB.62.12963}}.
\newline\urlprefix\url{https://link.aps.org/doi/10.1103/PhysRevB.62.12963}

\bibitem{1999Petterson_JAP}
L.~A. Pettersson, L.~S. Roman, O.~Ingan{\"a}s, Modeling photocurrent action
  spectra of photovoltaic devices based on organic thin films, JAP 86~(1)
  (1999) 487--496.

\bibitem{SOPRA}
\href{http://sspectra.com/sopra.html}{Sopra database}.
\newline\urlprefix\url{http://sspectra.com/sopra.html}

\bibitem{1988Gee_PIEEE}
J.~M. Gee, The effect of parasitic absorption losses on light trapping in thin
  silicon solar cells, in: Photovoltaic Specialists Conference, 1988.,
  Conference Record of the Twentieth IEEE, 1988, pp. 549--554 vol.1.
\newblock \href {http://dx.doi.org/10.1109/PVSC.1988.105762}
  {\path{doi:10.1109/PVSC.1988.105762}}.

\bibitem{2002Green_PPRA}
M.~A. Green, Lambertian light trapping in textured solar cells and
  light-emitting diodes: analytical solutions, Progress in Photovoltaics:
  Research and Applications 10~(4) (2002) 235--241.

\bibitem{1982Yablo_TED}
E.~Yablonovitch, G.~D. Cody, Intensity enhancement in textured optical sheets
  for solar cells, Electron Devices, IEEE Transactions on 29~(2) (1982)
  300--305.

\bibitem{GD_Calc}
\href{http://www.kjinnovation.com}{The grating diffraction calculator
  (gd-calc)}.
\newline\urlprefix\url{http://www.kjinnovation.com}

\bibitem{2014Lumb_JAP}
M.~P. Lumb, M.~A. Steiner, J.~F. Geisz, R.~J. Walters, Incorporating photon
  recycling into the analytical drift-diffusion model of high efficiency solar
  cells, Journal of Applied Physics 116~(19) (2014) 194504.

\bibitem{2012Jolley_PPRA}
G.~Jolley, L.~Fu, H.~Lu, H.~H. Tan, C.~Jagadish, The role of intersubband
  optical transitions on the electrical properties of ingaas/gaas quantum dot
  solar cells, Prog. Photovolt: Res. Appl.\href
  {http://dx.doi.org/10.1002/pip.2161} {\path{doi:10.1002/pip.2161}}.

\bibitem{2011King_PPRA}
R.~King, D.~Bhusari, A.~Boca, D.~Larrabee, X.-Q. Liu, W.~Hong, C.~Fetzer,
  D.~Law, N.~Karam, Band gap-voltage offset and energy production in
  next-generation multijunction solar cells, Progress in Photovoltaics:
  Research and Applications 19~(7) (2011) 797--812.

\bibitem{2017Cappelluti_ESPC}
{Cappelluti, F.}, {Ghione, G.}, {Gioannini, M.}, {Bauhuis, G.}, {Mulder, P.},
  {Schermer, J.}, {Cimino, M.}, {Gervasio, G.}, {Bissels, G.}, {Katsia, E.},
  {Aho, T.}, {Niemi, T.}, {Guina, M.}, {Kim, D.}, {Wu, J.}, {Liu, H.},
  \href{https://doi.org/10.1051/e3sconf/20171603007}{Novel concepts for
  high-efficiency lightweight space solar cells}, E3S Web Conf. 16 (2017)
  03007.
\newblock \href {http://dx.doi.org/10.1051/e3sconf/20171603007}
  {\path{doi:10.1051/e3sconf/20171603007}}.
\newline\urlprefix\url{https://doi.org/10.1051/e3sconf/20171603007}

\bibitem{2010Tommila_SEMSC}
J.~Tommila, V.~Poloj{\"a}rvi, A.~Aho, A.~Tukiainen, J.~Viheri{\"a}l{\"a},
  J.~Salmi, A.~Schramm, J.~Kontio, A.~Turtiainen, T.~Niemi, et~al.,
  Nanostructured broadband antireflection coatings on alinp fabricated by
  nanoimprint lithography, Solar Energy Materials and Solar Cells 94~(10)
  (2010) 1845--1848.

\bibitem{2016Musu_OSA}
A.~Musu, F.~Cappelluti, T.~Aho, V.~Poloj{\"a}rvi, T.~K. Niemi, M.~Guina,
  Nanostructures for light management in thin-film gaas quantum dot solar
  cells, in: Solid-State Lighting, Optical Society of America, 2016, pp.
  JW4A--45.

\end{thebibliography}

\end{document}